\documentclass[preprintnumbers,twocolumn,floats,nofootinbib,superscriptaddress,aps,balancelastpage]{revtex4}
\usepackage{graphicx}
\usepackage{amsmath,amssymb,amsfonts}
\usepackage{hyperref}
\usepackage{slashed}
\newcommand{\nur}{\tilde{n}} 
\newcommand{\tnu}{\tilde{\nu}_1} 
\newcommand{\nuchi}{\tilde{\chi}_1^0} 
\newcommand{\nue}{\tilde{\chi}_1^0} 
\newcommand{\Neff}{N_{\rm{eff}}}

\newcommand{\PArthENoPE}{\texttt{PArthENoPE}}

\begin{document}
\title{A lower bound on the mass of cold thermal dark matter from Planck
}

\preprint{IPPP/13/18, DCPT/13/36}

\author{C\'eline B\oe hm}
  \affiliation{Institute for Particle Physics Phenomenology, Durham University, South Road, Durham, DH1 3LE, United Kingdom}
\affiliation{LAPTH, U. de Savoie, CNRS,  BP 110,
  74941 Annecy-Le-Vieux, France\\
{\smallskip \tt  \footnotesize \href{mailto:c.m.boehm@durham.ac.uk}{c.m.boehm@durham.ac.uk}, \href{mailto:m.j.dolan@durham.ac.uk}{m.j.dolan@durham.ac.uk}, \href{mailto:christopher.mccabe@durham.ac.uk}{christopher.mccabe@durham.ac.uk}} \smallskip}
\author{Matthew J.~Dolan}
\author{Christopher McCabe}
\affiliation{Institute for Particle Physics Phenomenology, Durham University, South Road, Durham, DH1 3LE, United Kingdom}

\begin{abstract} 
We show that the new measurement of the effective number of neutrinos ($\Neff$) by the Planck satellite can be used to set a robust lower bound on the mass of cold thermal dark matter of $\mathcal{O}(\rm{MeV})$. Our limit applies if the dark matter remains in thermal equilibrium by coupling to electrons and photons or through interactions with neutrinos, and applies regardless of whether the dark matter annihilation cross-section is $s$-wave or $p$-wave. To illustrate our bounds we apply them to a model of a supersymmetric neutralino annihilating to neutrinos, via a light mixed left-right handed sneutrino mediator. While this scenario was not constrained by previous data, the Planck limits on $\Neff$ allow us to set a lower bound on the neutralino dark matter mass of 3.5~MeV.
 \end{abstract} 

\maketitle

\section{Introduction}
The mass and non-gravitational interactions of particle dark matter remain unknown. Although many candidates have been proposed, particles that achieve the observed relic abundance through thermal freeze out have garnered the most attention~\cite{Zeldovich1,Zeldovich2,Chiu1}. 
While thermal relics are often assumed to have weak-scale masses this does not have to be the case; the thermal freeze-out mechanism works for a much broader range of masses and leaves room for models from keV to multi-TeV masses encompassing the whole experimentally allowed range of cold dark matter.

For thermal relics, an upper bound of approximately $340$~TeV can be set on the dark matter mass using partial wave-unitarity~\cite{Griest:1989wd}. Finding a robust lower bound is more difficult. General theoretical attempts have been made, for example by Hut and Lee-Weinberg~\cite{Hut:1977zn,Lee:1977ua}. However, these can be avoided by keeping the dark matter particle in thermal equilibrium with, for instance, a light mediator~\cite{Boehm:2003hm}, and a robust lower bound requires using experimental constraints. For example, a model independent lower bound for fermionic dark matter of $m\gtrsim1$~keV can be set by considering the phase-space distribution of dark matter in dwarf spheroidal galaxies~\cite{Boyarsky:2008ju}. Structure formation indicates that dark matter cannot be hot so a similar bound of $\mathcal{O}(\rm{keV})$ can be placed by requiring that the free-streaming length is small~\cite{Bond:1983hb}.
A more model dependent lower bound, $m\gtrsim10$~GeV, can be placed by considering distortions due to energy injection by \mbox{$s$-wave} annihilation in the cosmic microwave background (CMB)~\cite{Padmanabhan:2005es,Mapelli:2006ej,Zhang:2006fr,Galli:2009zc,Slatyer:2009yq,Hutsi:2011vx,Galli:2011rz,Finkbeiner:2011dx,Slatyer:2012yq,Lopez-Honorez:2013cua}. However this can be evaded if the annihilation is a $p$-wave process or if the dark matter annihilates to neutrinos. Bounds can also be placed from Big-Bang Nucleosynthesis~\cite{Kolb:1986nf,Serpico:2004nm}, but these suffer from systematic uncertainties in the measured values of the primordial nuclear abundances~\cite{Beringer:1900zz,2012arXiv1208.0032S}.

Here we find a new lower bound for cold thermal dark matter using data from the Planck satellite. The only light states in the Standard Model (SM) that it is possible to  be in thermal equilibrium with are the neutrinos or electrons and photons, and so it is natural to assume that the dark matter will be coupled to these states. If this is the case and the dark matter decouples while non-relativistic, there will be a change in the effective number of neutrinos ($\Neff$) through its effect on the neutrino-photon temperature ratio~\cite{Boehm:2012gr,Ho:2012ug,2013arXiv1303.0049S}.
Using the results from Planck~\cite{Ade:2013lta}, which has measured the CMB angular power spectrum with an unprecedented accuracy, we show that a bound of about an MeV can be placed on the dark matter mass, if the dark matter is in thermal equilibrium with the neutrinos or the electrons and photons after neutrino decoupling. The exact bound depends on whether the dark matter is a scalar or fermion and to which SM species it couples. The advantage of our method is that it works even for a $p$-wave annihilation cross-section and for annihilation to neutrinos - cases when limits such as energy injection into the CMB do not apply.  Previous studies have used $\Neff$ to constrain the mass of dark matter annihilating exclusively into electrons~\cite{Ho:2012ug}. However, that analysis did not take into account the degeneracy of $\Neff$ with helium, which leads to a slightly weaker bound, while ours does. Our analysis using the Planck data supersedes that study and we also consider the case of annihilation into neutrinos. 

Our bounds apply to models such as electric dipole and anapole dark matter~\cite{Bagnasco:1993st, Pospelov:2000bq,Sigurdson:2004zp,Masso:2009mu, An:2010kc, Fitzpatrick:2010br,2012arXiv1211.0503H} or to models in~\cite{Boehm:2003bt,Ahn:2005ck,Hooper:2008im} where the dark matter annihilates into electrons or photons, and could be relevant to models which give a signal at electron recoil direct detection experiments~\cite{Essig:2011nj,Essig:2012yx,Graham:2012su}. Fewer models have been constructed where dark matter annihilates into neutrinos. Therefore, as an application of our bounds, we constrain a supersymmetric dark matter model where the dark matter is a light bino-like neutralino and a mixed left-right handed sneutrino plays the role of the light mediator. Annihilation in this model is $p$-wave and the neutralino remains in equilibrium with the neutrinos. We show that a new lower bound can be placed on the neutralino mass of~$3.5$~MeV.

This paper is organised as follows. In section~\ref{sec:lowerneff} we show how light, cold, thermal dark matter annihilating into neutrinos or electrons and photons changes $\Neff$ and derive a bound on the mass of dark matter particle using data from Planck. We compare these bounds with those derived from Big Bang Nucleosynthesis (BBN) and show that they are stronger. Following that, in section~\ref{sec:lightchi} we construct a supersymmetric model where the dark matter is a light bino-like neutralino and a light mixed left-right sneutrino mediates efficient neutralino annihilation into neutrinos.  Appendices discuss changes in the mass bounds when the HST prior on $H_0$ is applied and the collected constraints on our neutralino model from invisible widths of the $Z$- and Higgs boson, rare meson decays, collider direct production and beam dump searches as well as constraints from SN1987A, large scale structure and direct detection experiments.

\section{A lower mass bound from Planck}
\label{sec:lowerneff}
In this section we show how a lower bound can be placed on the mass of cold thermal dark matter by using Planck's constraint on $\Neff$. This bound applies if the dark matter remains in thermal equilibrium with either photons and electrons or neutrinos until after the time when neutrinos decouple from electrons.\footnote{Our bounds do not apply to dark matter candidates that were not in thermal equilibrium with neutrinos, electrons or photons in the early universe. This includes, for instance, primordial black holes~\cite{Ricotti:2007au,Frampton:2010sw,Hawkins:2011qz} and the QCD axion, whose abundance arises from the non-thermal misalignment mechanism~\cite{Preskill:1982cy,Abbott:1982af,Dine:1982ah}. However, non-QCD (heavier) axions can be in thermal equilibrium and can change $\Neff$ (see e.g.~\cite{Cadamuro:2010cz}).}

The effective number of neutrinos, $\Neff$, is a convenient way to parameterise the energy density apart from the photon contribution. It allows us to write the total energy density as
\begin{equation}\label{eq:rhotot}
\rho\equiv\rho_{\gamma}\left[1+\frac{7}{8} \left(\frac{4}{11}\right)^{4/3} \Neff \right]\;.
\end{equation}
The factor $(4/11)^{4/3}$ is the fourth power of the neutrino-photon temperature ratio $T^0_{\nu}/T_{\gamma}=(4/11)^{1/3}$, which assumes instantaneous decoupling of three neutrino species. In the standard cosmological model \mbox{$\Neff=3.046$}, reflecting that the actual neutrino-photon temperature relation is slightly higher due to decoupling effects~\cite{Mangano:2001iu, Mangano:2005cc}.

The standard lore is that a value of $\Neff>3.046$ implies that there must be a new light species that is relativistic when the cosmic microwave photons decouple (at $T\sim1$~eV).\footnote{See refs.~\cite{Brust:2013ova,DiBari:2013dna} for constraints on this class of models using Planck data.} However, this is not the case and thermal dark matter with mass up to $\sim10$~MeV can also change $\Neff$, as we now argue using the example of electrons. In the standard cosmological model, the electrons do not directly contribute to the energy density at recombination because they are non-relativistic at this time. However they do have an effect on $\Neff$ because the electrons caused the photons to develop a higher temperature than the neutrinos. The definition of $\Neff$ assumes that a definite temperature ratio holds, namely that \mbox{$T^0_{\nu}/T_{\gamma}=(4/11)^{1/3}$}. Had the electrons not been present, the photons and neutrinos would have had the same temperature and $\Neff$ would have been larger than three by a factor $(11/4)^{4/3}$.

In a similar fashion, thermal dark matter in thermal equilibrium with photons, electrons or neutrinos can change $\Neff$ by altering the neutrino-photon temperature ratio. In order for this to happen, the dark matter must transfer its entropy to either the electrons and photons or to the neutrinos, after the neutrinos decouple from the electrons at $T_{\rm{D}}\approx2.3$~MeV~\cite{Enqvist:1991gx} (we assume that new interactions do not change the neutrino decoupling temperature). If the dark matter is in thermal equilibrium with photons, electrons and neutrinos after neutrino decoupling, it will reheat them equally and $\Neff$ will not change. Since the entropy transfer occurs when the particle becomes non-relativistic, a change in $\Neff$ occurs if the mass $m\lesssim\text{few}\times T_{\rm{D}}\sim10$~MeV. Cold thermal dark matter automatically satisfies the condition of being in thermal equilibrium until the temperature drops below its mass; typically, MeV-mass thermal dark matter drops out of chemical equilibrium when $T\sim m/15$.\footnote{Our bounds also apply to a sub-component of cold dark matter that is a thermal relic; a thermal sub-component (with MeV-mass) remains in chemical equilibrium until $T\lesssim m/15$.}

\begin{figure*}[t!]
\includegraphics[width=2.05\columnwidth]{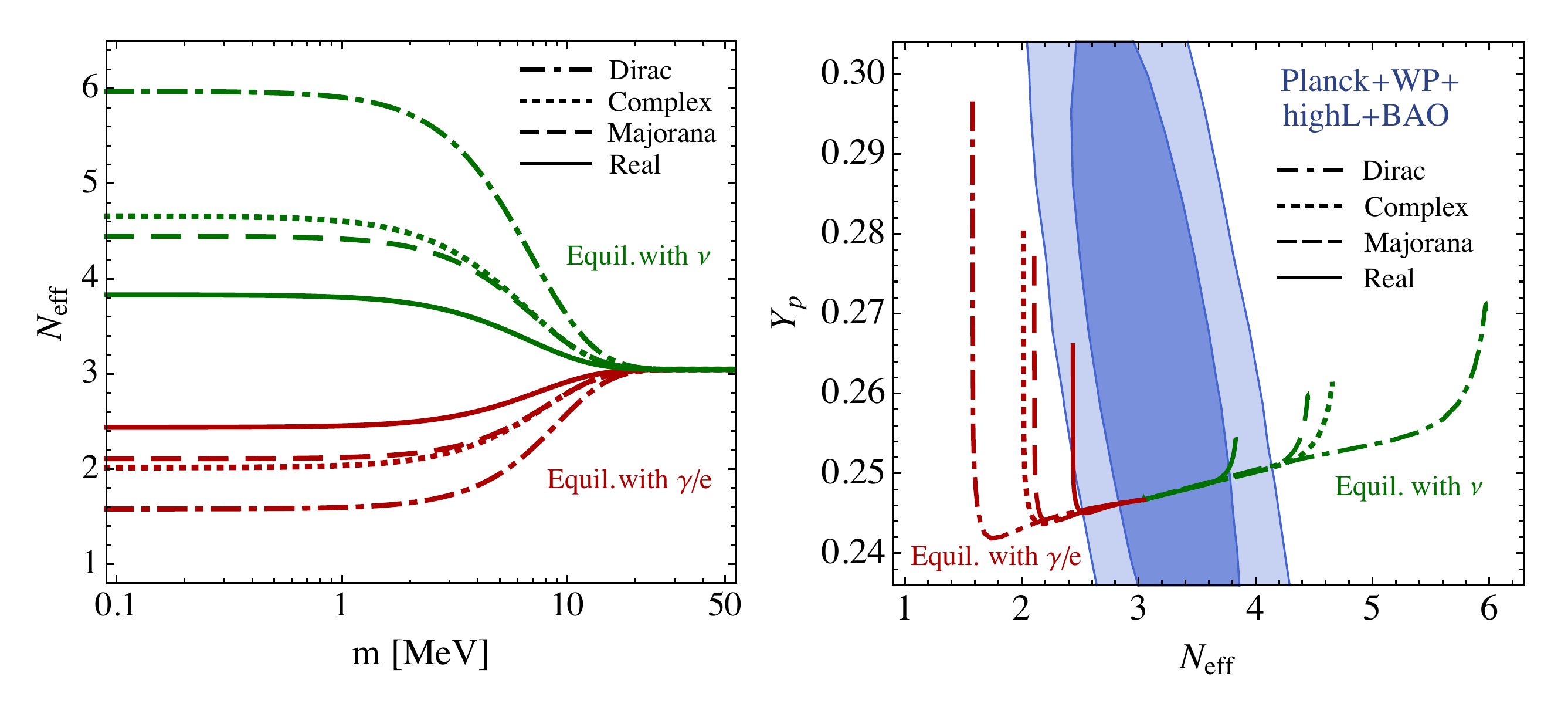}
\caption{Left panel: $\Neff$ as a function of the cold thermal dark matter mass $m$. The green (red) lines are for the case when the dark matter is in thermal equilibrium with neutrinos (electrons and photons) and show that $\Neff$ increases (decreases) as $m$ is reduced. Right panel: The blue regions show the $68\%$ and $95\%$ regions determined from Planck+WP+highL+BAO when both $\Neff$ and $Y_p$ are varied freely. The green (red) lines indicate the relationship between $Y_p$ and $\Neff$ for particles in thermal equilibrium with neutrinos (electrons and photons). As $m$ decreases, the prediction for $\Neff$ and $Y_p$ falls outside of the Planck confidence regions.}
\label{fig:mNeff}
\end{figure*}

We now put the above argument on a more quantitative footing. We first consider the case where the dark matter is in thermal equilibrium with neutrinos. The case where it is in equilibrium with electrons and photons then follows in a straightforward fashion. The contribution of $N_{\nu}$ neutrinos and $n$ particles (with mass $m_i$ and internal degrees of freedom $g_i$) in thermal equilibrium with the neutrinos to the energy density is
\begin{equation}\label{eq:rhonu}
\frac{\rho_{\nu+n}}{\rho_{\gamma}}=\frac{7}{8}\left(\frac{T_{\nu}}{T_{\gamma}}\right)^4 \left[N_{\nu}+\sum_{i=1}^n\frac{g_i}{2} I \left( \frac{m_i}{T_{\nu}}\right) \right]\;,
\end{equation}
where
\begin{align}
\label{eq:Iy}
I(x)=\frac{120}{7 \pi^4} \int^{\infty}_x dy \frac{y^2 \sqrt{y^2-x^2}}{e^{y}\pm1}
\end{align}
with limits $I(\infty)=0$ and $I(0)=1(8/7)$ for fermions (bosons) respectively. As usual, the sign $+\,(-)$ refers to fermion (boson) statistics. In general, $T_{\nu}\neq T_{\nu}^0$. We allow for $n$ particles as, in addition to the cold thermal dark matter, there may be light mediators that also contribute to the energy density. Comparing eqs.~\eqref{eq:rhotot} and~\eqref{eq:rhonu}, we see that
\begin{equation}
N_{\rm{eff}}=\left(\frac{4}{11}\right)^{-4/3}\left(\frac{T_{\nu}}{T_{\gamma}}\right)^{4} \left[N_{\nu}+\sum_{i=1}^n\frac{g_i}{2} I \left( \frac{m_i}{T_{\nu}}\right)   \right]\;.
\label{eq:Neff1}
\end{equation}
Anticipating that the bound on $m_i$ is such that $m_i\gg T_{\nu}(\text{at recombination})\sim1$~eV, we set $I(m_i/T_{\nu})=0$ so that
\begin{equation}
N_{\rm{eff}}=N_{\nu} \left(\frac{4}{11}\right)^{-4/3}\left(\frac{T_{\nu}}{T_{\gamma}}\right)^{4}\;.
\label{eq:Neff2}
\end{equation}

The ratio $T_{\nu}/T_{\gamma}$ is determined by considering entropy conservation (see e.g.~\cite{Kolb:1986nf,Srednicki:1988ce,Boehm:2012gr}). After neutrino decoupling at $T_{\rm{D}}\approx 2.3$~MeV, the entropy of the `neutrino plasma' and `electromagnetic plasma' are separately conserved so that (for $T_{\gamma}<T_{\rm{D}}$)
\begin{equation}\label{eq:Tnugamma}
\frac{T_{\nu}}{T_{\gamma}}=\left(\left. \frac{g_{\star s: \nu}}{g_{\star s: \gamma}}\right|_{T_{\rm{D}}} \frac{g_{\star s: \gamma}}{g_{\star s: \nu}} \right)^{1/3}\;.
\end{equation}
Here $\left.\right|_{T_{\rm{D}}}$ indicates that $g_{\star s}$ should be evaluated at the neutrino decoupling temperature $T_{\rm{D}}$ while $g_{\star s: \nu}$ and $g_{\star s: \gamma}$, defined through $s_{\nu}=2 \pi^2 g_{\star s: \nu} T_{\nu}^3/45$ and $s_{\gamma}=2 \pi^2 g_{\star s: \gamma} T_{\gamma}^3/45$ respectively, are the effective number of relativistic degrees of freedom in the neutrino and electromagnetic plasmas. Explicitly, 
\begin{equation}
g_{\star s:\nu}= \frac{14}{8}\left[ N_{\nu}+\sum_{i=1}^n \frac{g_{i}}{2} F\left(\frac{m_i}{T_{\nu}}\right)\right]\;.
\label{gstarsnu}
\end{equation}
where
\begin{equation}
F(x)=\frac{30}{7 \pi^4}\int^{\infty}_x dy \frac{(4 y^2-x^2)\sqrt{y^2-x^2}}{e^{y}\pm1}\;.
\end{equation}
with limits $F(\infty)=0$ and $F(0)=1(8/7)$ for fermions (bosons) respectively and the sign $+\,(-)$ refers to fermion (boson) statistics. 

Again, anticipating that the bound on $m_i$ is such that $m_i\gg T_{\nu}(\text{at recombination})\sim1$~eV, we find that for particles only in thermal equilibrium with neutrinos, eq.~\eqref{eq:Neff2} simplifies to
\begin{equation}
\label{eq:Neff3}
N_{\rm{eff}}^{\mathrm{Equil.}\, \nu}=N_{\nu} \left[1+\frac{1}{N_{\nu}}\sum_{i=1}^n \frac{g_{i}}{2} F\left(\frac{m_i}{T_{\rm{D}}}\right)\right]^{4/3}
\end{equation}

For the case of particles in thermal equilibrium with electrons or photons, we again find eq.~\eqref{eq:Neff2} and can use eq.~\eqref{eq:Tnugamma} to find the new temperature ratio. In this case, we find 
\begin{equation}
\label{eq:Neff4}
N_{\rm{eff}}^{\rm{Equil.}\, \gamma/e}=N_{\nu} \left[ 1+\frac{7}{22}\sum_{i=1}^n \frac{g_i}{2} F\left( \frac{m_i}{T_{\rm{D}}}\right) \right]^{-4/3}
\end{equation}
where we have used $F\left(m_e/T_{\rm{D}}\right)\approx1$.

The dot-dashed, dashed, dotted and solid lines in the left panel of fig.~\ref{fig:mNeff} show the value of $\Neff$ for a single particle of mass $m$ for a Dirac fermion, Majorana fermion, complex scalar and real scalar respectively. The case where the particle is in equilibrium with neutrinos is shown by the green lines. Here, $\Neff$ increases above the standard value of $\Neff=3.046$ for particles lighter than~$\simeq 20$ MeV. Conversely, $\Neff$ decreases below the standard value for particles in equilibrium with electrons and photons, as indicated by the red lines. There is no effect above $m\geq20$~MeV because the entropy transfer occurs before the electromagnetic and neutrino plasmas decouple resulting in the standard neutrino-photon temperature ratio. 

With eqs.~\eqref{eq:Neff3} and~\eqref{eq:Neff4} we can put a bound on the dark matter mass by requiring that $\Neff$ is compatible with the measured value from Planck. The central result from~\cite{Ade:2013lta},
\begin{equation}\label{eq:Neffplanck}
\Neff=3.30^{+0.54}_{-0.51}\quad(\text{95\%; Planck+WP+highL+BAO}),
\end{equation}
which combines Planck precision measurement of the CMB, WMAP-9's polarisation data (WP)~\cite{Bennett:2012fp}, SPT's high-$\ell$ measurement (highL)~\cite{Reichardt:2011yv} and baryon acoustic oscillations (BAO) measurements from large scale structure surveys~\cite{Percival:2009xn,Padmanabhan:2012hf,Blake:2011en,Anderson:2012sa,Beutler:2011hx} cannot be used to set an accurate constraint on this scenario. This is because this result assumes that all of the relativistic components parameterised by $\Neff$ consist of free streaming relativistic particles which are effectively massless. In particular, the helium abundance $Y_p$ is fixed to the BBN theory prediction for effectively massless relativistic particles. As was first described in \cite{Kolb:1986nf}, this relation does not hold in our case: the new particles alter the standard prediction for $Y_p$ from relativistic particles because the additional semi-relativistic particles and the change in the neutrino-photon temperature ratio both contribute to the energy density during BBN. The difference in this relationship is important because the impact of $Y_p$ on the damping tail of the power spectrum is degenerate with the effect from $\Neff$ (see e.g.~\cite{Hou:2011ec} for further discussion).

Therefore, we use the results from the Planck+WP+highL+BAO analysis in which $Y_p$ and $\Neff$ are simultaneously constrained. The $68\%$ and $95\%$ confidence regions from this analysis are shown in the right panel of fig.~\ref{fig:mNeff}. The green and red lines overlying these regions are our calculations for $Y_p$ (to be discussed in detail in the next section) against $\Neff$ for the cases of particles in equilibrium with neutrinos and electrons or photons respectively. As the mass $m$ of these particles decreases, the prediction for $\Neff$ eventually falls outside the Planck $95\%$ confidence region. Particles lighter than this are excluded at the $95\%$ confidence level.

Requiring that $\Neff$ is consistent with the Planck result, we exclude the following particle masses at $95\%$~C.L.\ for cold thermal dark matter particles in equilibrium with neutrinos:
\begin{align}
&\text{Real scalar}\;&\text{No constraint} \label{eq:nu_RScalarBound}\\
&\text{Complex scalar}\;&m<3.9~\text{MeV}\\ \label{eq:MajFermBound}
&\text{Majorana fermion}\;&m<3.5~\text{MeV}\\ 
&\text{Dirac fermion}\;&m<7.3~\text{MeV}\label{eq:nu_DirFermBound}
\end{align}
Similarly, we can exclude the following cold thermal dark matter particle masses at $95\%$~C.L.\ when in equilibrium with electrons and photons:
\begin{align}
&\text{Real scalar}\;&0.4~\text{MeV}<m<2.6~\text{MeV}\label{eq:e_RScalarBound}\\
&\text{Complex scalar}\;&m<6.5~\text{MeV}\\
&\text{Majorana fermion}\;&m<6.4~\text{MeV}\\
&\text{Dirac fermion}\;&m<9.4~\text{MeV}\label{eq:e_DirFermBound}
\end{align}

Had we set the limits with eq.~\eqref{eq:Neffplanck}, the bounds would be about $30\%$ higher. These bounds are independent of whether the annihilation cross-section is $s$- or $p$-wave, but there are ways to evade them. We mention three: firstly, the dark matter's abundance might arise from a non-thermal mechanism so that agreement with the Planck limit is found. Secondly, while dark matter in thermal equilibrium with electrons and photons would by itself decrease $\Neff$, the presence of some extra relativistic degrees of freedom (`dark radiation' or sterile neutrinos) would increase $\Neff$ bringing it back into agreement with the Planck result~\cite{2012arXiv1212.1689H,2013arXiv1303.0049S,Drewes:2013gca}. Thirdly, the dark matter could be in thermal equilibrium with electrons, neutrinos and photons in which case there is no change in the standard neutrino-photon temperature ratio and hence, no change in $\Neff$. However, the BBN bounds on the second scenario will be more stringent since the dark matter and the extra degrees of freedom both increase $Y_p$ and there are strong constraints from $\nu-e$ scattering (see e.g.~\cite{Auerbach:2001wg, Formaggio:2013kya}) on the third scenario.

Finally, although we considered here the degeneracy of $Y_p$ and $\Neff$, it is also the case that $\Neff$ is positively correlated with the Hubble constant $H_0$. The Planck measurement of $H_0$ is about $2.5 \sigma$ lower than direct astrophysical measurements. When the astrophysical measurements of $H_0$ are used as a prior, slightly higher values of $\Neff$ are preferred. For completeness, we give the resulting mass bounds in appendix~\ref{app:H0}.

\subsection{Comparison with Big Bang Nucleosynthesis}
\label{app:BBN}
\begin{figure*}[t!]
\includegraphics[width=2.05\columnwidth]{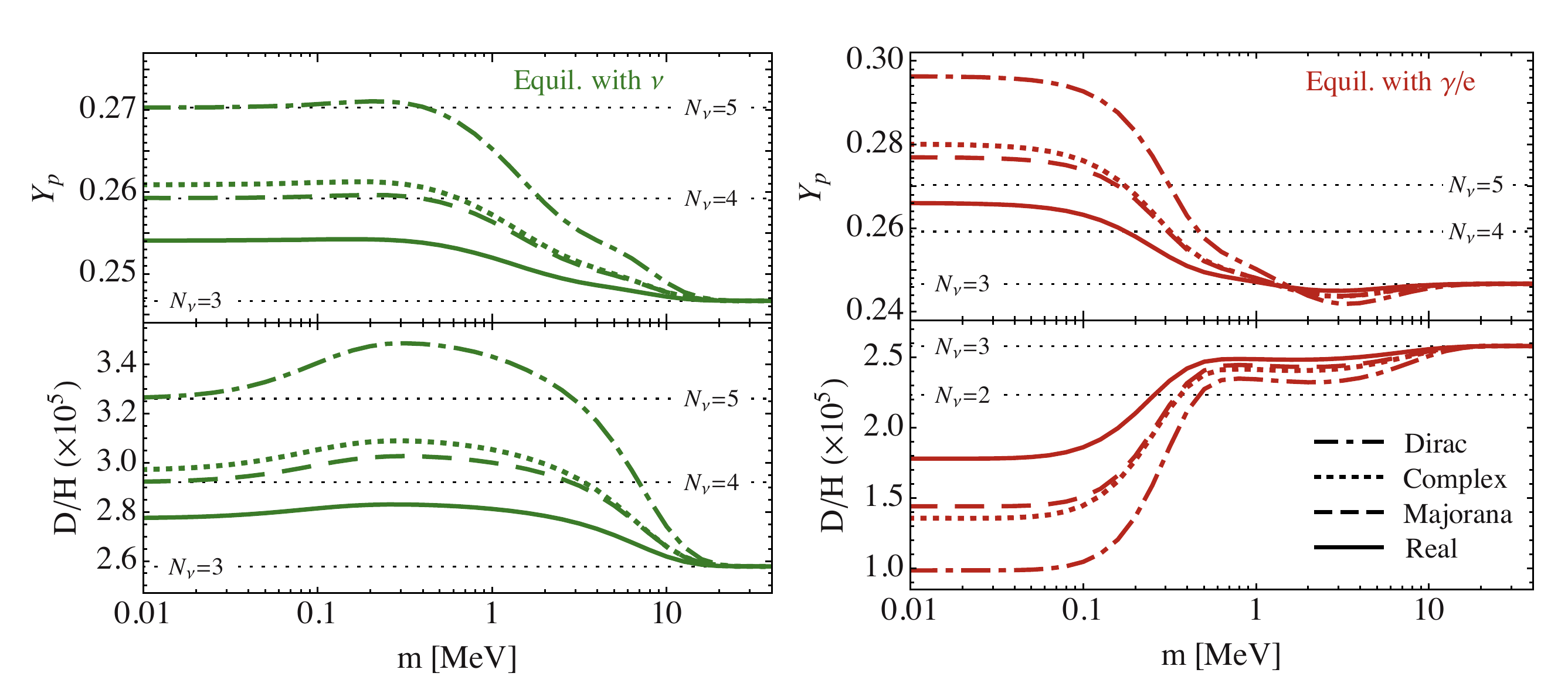}
\caption{ The left [right] panels show the calculated abundances of helium (upper segment) and deuterium (lower segment) for cold dark matter in thermal equilibrium with neutrinos [photons and electrons] respectively. The dotted lines label by $N_{\nu}$ show the predicted abundances for $N_{\nu}$ massless neutrinos.}
\label{fig:BBN}
\end{figure*}

A bound on the dark matter mass can also be placed by comparing the primordial abundance of helium and deuterium predicted from BBN with the observed values. Here we compare those bounds with those we have just derived from Planck's measurement of $\Neff$. Unfortunately, the observed values of primordial helium and deuterium are beset by large systematic errors and various central values and errors appear in the literature. Following the convention in the rest of particle physics, we take the values recommended in the PDG review of BBN~\cite{Beringer:1900zz}: 
\begin{align}
\label{PDG:BBN}
Y_p&=0.249 \pm 0.009\\
\rm{D}/\rm{H}&=(2.81 \pm 0.21) \times 10^{-5}\;.
\end{align}

We calculate $Y_p$ and $\rm{D}/\rm{H}$ as a function of particle mass $m$ with a modified version of the $\PArthENoPE$ BBN code~\cite{Pisanti:2007hk}. As well as including the energy density of the additional particle, the effects of the change in the neutrino-photon temperature relation must also be accounted for. This includes, for instance, a modification of the interaction rates for $p \leftrightarrow n$ which change the ratio of the neutron-to-proton number density. We refer the reader to~\cite{Serpico:2004nm,Boehm:2012gr} for a detailed description of how these effects are implemented.

\begin{table}[t!]
\centering
\begin{tabular}{l c c c}
\hline
\multicolumn{4}{c}{Equilibrium with $\nu$} \\
	 & Planck & $Y_p$ \;\;&D/H  \\[0.3ex]
\hline \\[-1.0em]
 Real scalar 		& -- & -- \;\;& --  \\
Complex scalar 	& 3.9  & -- \;\;& -- \\
Majorana fermion	& 3.5 & -- \;\; & --  \\
Dirac fermion		& 7.3 & 0.8 \;\;& 3.3 \\[0.5ex]
\hline
\multicolumn{4}{c}{Equilibrium with $\gamma / e$} \\
	 & Planck & $Y_p$ \;\;&D/H  \\[0.5ex]
\hline  \\[-1.0em]

 Real scalar 		& $0.4 \leq m \leq 2.6$ & -- \;\;& 0.4 \\
Complex scalar 	& 6.5  & 0.2 \;\;& 0.5 \\
Majorana fermion	& 6.4  & 0.2 \;\;& 0.5 \\
Dirac fermion		& 9.4  &0.3 \;\; & 5.3 \\[0.5ex]
\hline
\end{tabular}
\caption{The upper (lower) table show the 95\% C.L.~bounds on the dark matter mass $m$ for particles in equilibrium with neutrinos (electrons and photons). The numbers are an upper bound unless stated otherwise, `--' indicates that there is no limit  and the units are MeV. Generally, the limit from D/H is stronger than the limit from $Y_p$, while Planck is stronger than both.}
\label{tab:bounds}
\end{table}

The abundances for $Y_p$ and D/H are shown in the upper and lower segments of fig.~\ref{fig:BBN} respectively. In calculating these values, we have taken $\Omega_b h^2=0.0224$ and $\tau_n=880.1$~s, as recommended by the PDG~\cite{Beringer:1900zz}. The left and right panels show the results for particles in thermal equilibrium with neutrinos and electrons or photons respectively. The dotted lines labelled by $N_{\nu}$ show the predicted abundances for $N_{\nu}$ massless neutrinos.

The resulting masses that can be excluded at the $95\%$~C.L.\ are given in table~\ref{tab:bounds}. Numbers quoted are in units of MeV. The bounds from D/H are generally stronger than those from $Y_p$ but both are weaker than the Planck bounds. The only exception is for a real scalar in thermal equilibrium with photons and electrons. With the combination of the D/H and Planck bounds, the region $m<2.6$~MeV is excluded.

\section{Application: light neutralino dark matter}
\label{sec:lightchi}
We now consider the application of the above bounds to a specific cold thermal dark matter candidate. In our model, the dark matter candidate is a light bino-like neutralino which remains in thermal equilibrium with neutrinos through stronger than weak interactions
mediated by a light mixed left-right handed sneutrino. The annihilation cross-section is $p$-wave suppressed and the above bound provides the strongest constraint on the mass. 

The natural expectation for the mass of the lightest neutralino is about the weak scale, $\mathcal{O}(100)$~GeV. However the absence of any signal of supersymmetry (SUSY) at the Large Hadron Collider (LHC) may indicate that the much-studied usual mechanisms of SUSY breaking are not realised, and/or that some tuning may be required. Yet SUSY with $R$-parity still has many desirable features, such as gauge coupling unification, and may still tame most of the electroweak hierarchy problem. It also provides a flexible framework that is extremely useful to explore possibilities for the dark matter parameter space. In light of these arguments, we consider an extension of the MSSM with a spectrum which is dramatically different from those considered in standard SUSY model building (summarised in fig.~\ref{fig:spectrum}).  

In the MSSM, the lightest neutralino is a superposition of bino ($\tilde{B}$), neutral wino ($\tilde{W}^0$) and neutral higgsinos ($\tilde{h}_d^0$ and $\tilde{h}_u^0$):
\begin{equation}
\nuchi=N_{11} \tilde{B}+N_{12} \tilde{W}^0+N_{13} \tilde{h}_d^0+N_{14} \tilde{h}_u^0\;,
\end{equation}
where $N_{1i}$ are the components of the neutralino mixing matrix~\cite{Dreiner:2008tw}. Searches at LEP impose a lower bound on $M_2$, which, when the GUT relation $M_1=5 \tan^2\theta_W M_2/3\approx M_2/2$ is imposed, translate into a lower bound on $M_1$; in this case a lower bound of 46~GeV can be placed on the mass of the lightest neutralino~\cite{Abdallah:2003xe,Beringer:1900zz}. However, when this assumption is relaxed and $M_1$ and $M_2$ are treated as independent parameters, a massless neutralino is not ruled out by laboratory, astrophysical or cosmological limits if it is bino-like~\cite{Bartl:1989ms, Choudhury:1999tn,Kachelriess:2000dz, Gogoladze:2002xp,Dreiner:2003wh, Dreiner:2009ic,Dreiner:2009er}. Therefore, here we consider the case when the dark matter is, for all purposes, purely bino-like.

In this paper we remain agnostic as to the exact mechanism that gives rise to a light neutralino. However, we briefly mention two mechanisms by which this could arise. First, a light bino-like neutralino can be obtained if there is a hierarchy between $M_1$ and $(M_2, \mu)$. This arises, for instance, in hybrid models of SUSY breaking~\cite{Dudas:2008eq}. In this case, in the limit $M_2, \mu \gg m_Z, M_1$, at tree-level the lightest neutralino mass is given by
\begin{equation}
m_{\nuchi}\approx N_{11}^2 \left|M_1 - \sin(2\beta)\sin^2 \theta_W \frac{m_Z^2}{\mu}\right| 
\end{equation}
and the bino-component is 
\begin{align}
|N_{11}&|\approx\left[1+\sin^2\theta_W \left( \frac{m_Z}{\mu} \right)^2 \right]^{-1/2}  \\
&\approx 1-0.01\left(\frac{300\text{ GeV}}{\mu} \right)^2\;.
\end{align}
The higgsino components are sub-dominant: \mbox{$|N_{13}|\approx \sin \beta \sqrt{1-N_{11}^2}$} and \mbox{$|N_{14}|\approx \cos \beta \sqrt{1-N_{11}^2}$} respectively, while the wino-component is negligible: \mbox{$N_{12}\approx\mathcal{O}(N_{11} M_1/M_2)$}.

Second, we can invoke a symmetry argument. In SUSY models with an $R$-symmetry, gaugino soft-masses are forbidden (Dirac masses are required for the wino and gluino but the bino does not require them). However, it is expected that supergravity effects violate the (global) $R$-symmetry giving rise to a light bino-like neutralino, such as in~\cite{Davies:2011mp}. In that article the Majorana neutralino mass is
\begin{equation}
M_1=\frac{11 \alpha}{4 \pi \cos^2 \theta_W}\,m_{3/2}\approx 9\times10^{-3}\,m_{3/2}\;.
\end{equation}
and the neutralino is totally bino-like. If the gravitino mass is low, $\mathcal{O}(100)$~GeV or less, then the lightest neutralino will be at the MeV scale.

\subsection{Thermal neutralino dark matter}
\label{sec:thermal}
\begin{figure}[t]
\includegraphics[width=0.99\columnwidth]{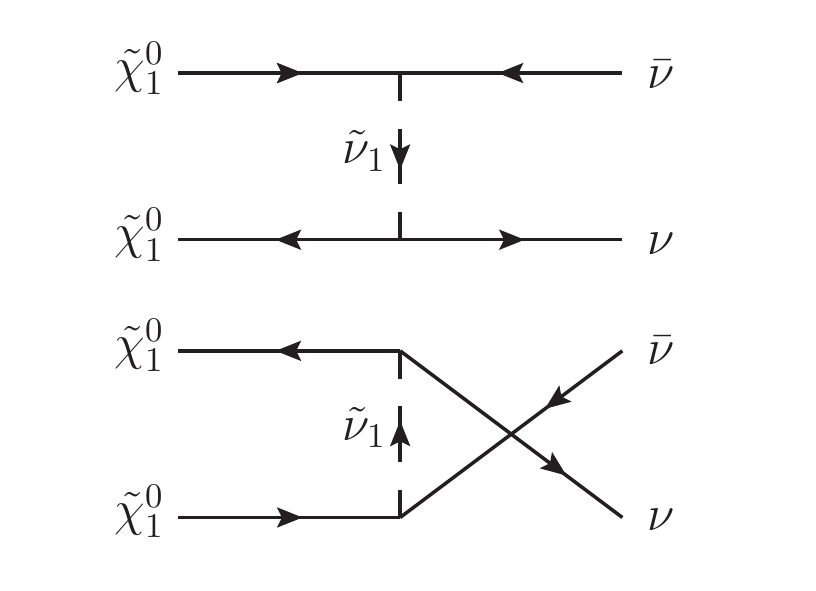}
\caption{Feynman diagrams showing the interaction that keep the neutralino in chemical equilibrium with the neutrinos in the early universe. The mediator $\tnu$ is a light mixed left-right handed sneutrino.}
\label{fig:Feynmandiagrams}
\end{figure}

The arguments of Hut~\cite{Hut:1977zn} and Lee-Weinberg~\cite{Lee:1977ua} suggest that a neutralino lighter than $\mathcal{O}(10)$~GeV cannot have the correct thermal relic abundance in the MSSM as all mediators with Standard Model particles are $\mathcal{O}(100)$~GeV. Indeed, detailed studies have borne this out~\cite{Hooper:2002nq, Bottino:2002ry, Belanger:2003wb}. We will therefore need to consider an extension of the MSSM which has a new light mediator in order to keep the neutralino in equilibrium with the thermal bath for longer.\footnote{Of course, this not the only option for a light neutralino to obtain the observed relic abundance: it could occur via a non-thermal mechanism or some source of entropy could dilute the neutralino overabundance.} The light neutralino can only annihilate to those states present in the early universe, namely electrons, photons and neutrinos. We choose as our mediator a light mixed left-right handed sneutrino, which keeps the neutralino in equilibrium with neutrinos through the diagrams shown in fig.~\ref{fig:Feynmandiagrams}. The mixing with a sterile right-handed component is necessary because the left-handed sneutrino couples strongly with the $Z$-boson. Constraints from the $Z$-width prevent it from being lighter than $m_Z/2$ (this is discussed in greater detail in appendix~\ref{app:particleconstraints}).

Therefore, in addition to the MSSM matter content we introduce three right-handed neutrino superfields $N_i$, which are gauge singlets with respect to the Standard Model gauge symmetries. In addition to the usual MSSM superpotential terms, we also include\begin{equation}
\Delta \mathcal{W}= Y^{\nu}_{ij}\, H_u \cdot L_i N_j \;,
\end{equation}
where $L_i$  is left-handed leptonic chiral superfield, $H_u$ is the up-type Higgs superfield and for $SU(2)$-doublets, our notation is $A\cdot B= A^2 B^1-A^1 B^2$. When the scalar component of $H_u$ gets a vev $(h_u=v \sin \beta)$, where $v=174$ GeV, this gives rise to a Dirac mass for the neutrinos. Therefore, the magnitude of the entries $Y$ are approximately $Y^{\nu}_{ij} \sim 10^{-12}$, and this matrix is diagonalised by the PMNS matrix~\cite{Pontecorvo:1957qd,Pontecorvo:1967fh,Maki:1962mu}. We do not consider a Majorana mass term in this article.\footnote{A Majorana neutrino mass term could be generated radiatively along the lines of~\cite{ArkaniHamed:2000bq, ArkaniHamed:2000kj,Borzumati:2000mc,Borzumati:2000ya, Arina:2007tm, MarchRussell:2004uf, MarchRussell:2009aq} by including a supersymmetric mass term $m_N NN$ and the accompanying soft $B$-term in the scalar potential. We do not investigate this further.}

With the additional soft terms, the left- and right-handed sneutrino contribution to the scalar potential is
\begin{equation}
V_{\rm{soft}}\supset m^2_{\tilde{\nu}_L} |\tilde{\nu}_{Li}|^2+ m^2_{\nur} |\nur_i|^2 + A_{ij}  h_u \cdot \tilde{L}_i \nur_j +\rm{h.c.}\;.
\end{equation}
For simplicity we have assumed that $m^2_{\tilde{\nu}_L}$ and $m^2_{\tilde{n}}$ are generation independent and that $A_{ij}$ has real entries with the same flavour structure as $Y^{\nu}_{ij}$. Therefore, $A_{ij}$ is also diagonalised by the PMNS matrix. This means that
after diagonalising $A_{ij}$, the sneutrino eigenstates $\tilde{\nu}^{\alpha}_{L}$ and $\tilde{n}^{\alpha}$ have the same flavour structure as the neutrinos. For instance, the neutrino mass eigenstate $\nu_2$ has approximately equal electron, muon and tau flavour so $\tilde{\nu}^2_{L}$ and $\tilde{n}^{2}$ also have approximately equal electron, muon and tau flavour.

The diagonal $A_{\alpha}$ terms induce mixing between the left- and right-handed eigenstates (without changing the flavour structure) so that the lightest mass eigenstate is
\begin{equation}
\tnu=-\sin\theta_1\; \tilde{\nu}^{\alpha}_{L}+\cos\theta_1\; \tilde{n}^{\alpha \star}
\end{equation}
with mass \begin{equation}
m^2_{\tnu}=\frac{m_{\tilde{n}}^2 \cos^2\theta_1-m_{\tilde{\nu}_{L}}^2 \sin^2\theta_1}{\cos 2\theta_1}\;.
\end{equation}
The mixing angle is given by
\begin{equation}\label{eq:mixingangle}
\tan 2\theta_i=\frac{2 A_i v \sin \beta}{m_{\tilde{\nu}_{L}}^2-m_{\tilde{n}}^2}\;.
\end{equation}
 For definiteness, we assume that the flavour index is $\alpha=2$ for the lightest mass eigenstate so that it couples to the electron, muon and tau neutrinos with equal strength (to a good approximation). In the remainder of the paper, we drop the subscript `1' on the mixing angle of the lightest mass eigenstate.
 
The left-handed sneutrino mass-term, $m_{\tilde{\nu}_{L}}=\tilde{m}_{L}^2+\frac{1}{2}\cos(2\beta) m_Z^2$, shares the soft-mass $\tilde{m}_{L}$ with the left-handed selectron so that the ATLAS bound \mbox{$m_{\tilde{e}_L}\gtrsim195$~GeV}~\cite{Aad:2012pxa} implies that $m_{\tilde{\nu}_{L}}\gtrsim200$~GeV. Therefore, to have a light sneutrino requires the relation $m_{\tilde{n}}\approx\tan\theta m_{\tilde{\nu}_{L}}$.  We assume that this cancellation occurs accidentally for the lightest sneutrino while the others remain at $\mathcal{O}(100)$~GeV. To achieve such a mass requires a tuning comparable to that required in the electroweak sector of the CMSSM~\cite{Buchmueller:2011ab,Buchmueller:2012hv}.

\begin{figure}[t]
\includegraphics[height=0.88\columnwidth]{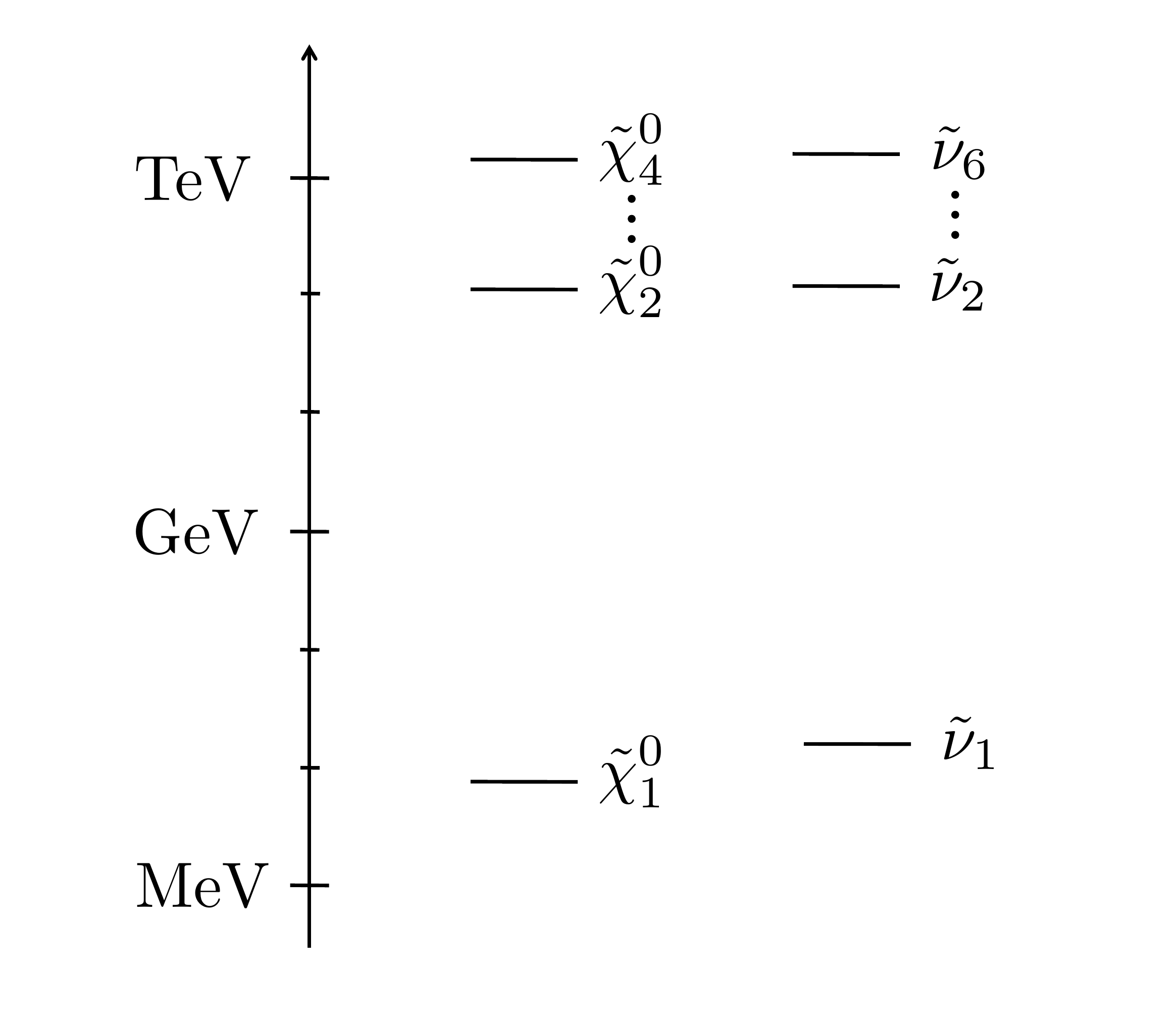}
\caption{A schematic diagram showing the structure of the neutralino and sneutrino sectors that we consider in this paper. As well as a sub-GeV bino-like neutralino, we have a light mixed left-right handed sneutrino. The other neutralinos and mixed sneutrinos have soft masses around the electroweak scale $\mathcal{O}(100)$~GeV. We assume that the other sparticles are heavy enough to have evaded direct search constraints from the LHC and LEP.}
\label{fig:spectrum}
\end{figure}

To summarise, the neutralino and sneutrino mass spectrum is shown in fig.~\ref{fig:spectrum}. In addition to the light bino-like neutralino, we also have a light mixed left-right handed sneutrino. We assume that all other sparticles, including the heavier sneutrinos and neutralinos are heavy enough to be consistent with direct search bounds from the LHC and LEP.  
 
\begin{figure}[t]
\includegraphics[height=0.88\columnwidth]{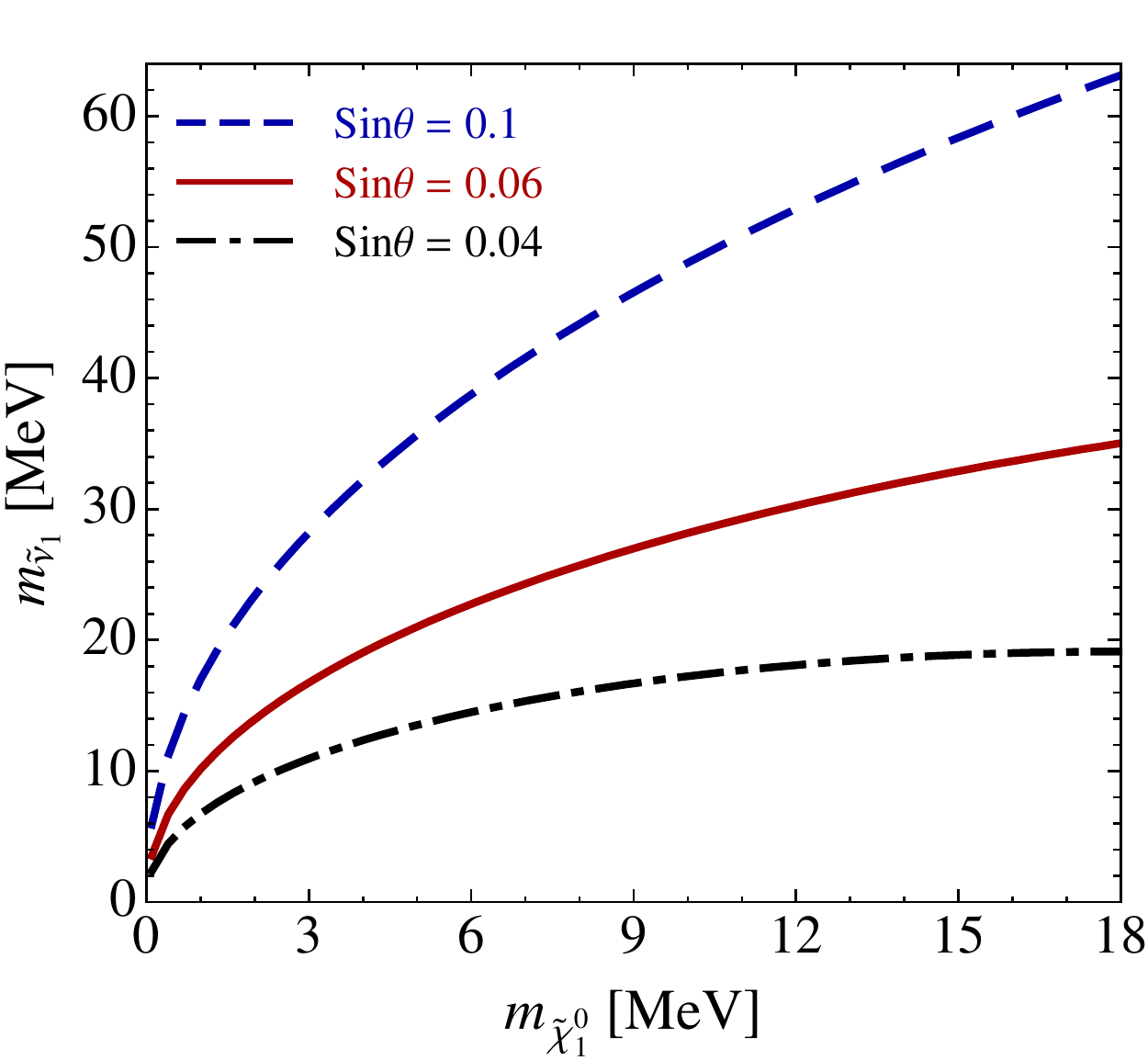}
\caption{The mixed sneutrino mass $m_{\tnu}$ required to achieve $\Omega_{\nue}h^2=0.11$ as a function of the neutralino mass $m_{\nue}$ for three different values of $\sin\theta$. The upper dashed blue line is for $\sin\theta=0.1$, the middle solid red line for $\sin\theta=0.06$, and the bottom dash-dotted black line for $\sin\theta=0.04$. The required sneutrino mass is typically $m_{\tnu}\sim\text{few}\times m_{\nue}$. }
\label{fig:relicabundance}
\end{figure}

We turn now to the calculation of the relic abundance. The dominant interaction contributing to neutralino annihilation with the neutrinos is the $t$- and $u$-channel exchange of the mixed sneutrino shown in fig.~\ref{fig:Feynmandiagrams}. In the non-relativistic limit, $t\approx u \approx-m_{\nue}^2$ and the resulting $p$-wave thermally averaged annihilation cross-section is
\begin{align}
\langle \sigma_{\rm{ann}} v \rangle &=\frac{G_F^2 m_Z^4 \sin^4\theta_W}{\pi}\frac{ |N_{11}|^4 \sin^4\theta}{(m_{\tnu}^2+m_{\nue}^2)^2} \frac{m_{\nuchi}^2}{x}\\
&\approx \frac{103 \text{ pb}}{x} \left(\frac{\sin\theta}{0.1}\right)^4 \left(\frac{m_{\nue}}{5\text{ MeV}}\right)^2 \left(\frac{35 \text{ MeV}}{m_{\tnu}}\right)^4\;,
\label{eq:sigmav}
\end{align}
where as usual, $x=m_{\nue}/T$. Using the well known analytic approximation for $p$-wave annihilation, for instance see~\cite{KolbTurner} for details, we find that the thermal relic abundance is
\begin{equation}
\Omega_{\nue}h^2=0.15 \left(\frac{s_0}{2889.23 \text{ cm}^{-3}} \right)\frac{\sqrt{g_{\star }|_{x_{\rm{f}}}} \left(x_{\rm{f}}/15\right)^{2} }{g_{\star s}|_{x_{\rm{f}}} \left(\tilde{\sigma}_0/100\, \rm{pb}\right)}\;,
\end{equation}
where $\tilde{\sigma}_0$ is defined through $\langle \sigma_{\rm{ann}} v \rangle = \tilde{\sigma}_0 x^{-1}$ and
\begin{equation}
\begin{split}
x_{\rm{f}}&=\ln \left[0.076 g g_{\star}^{-1/2}m_{\rm{pl}}m_{\nue}\tilde{\sigma}_0\right]\\
&\qquad-\frac{3}{2}\ln \left\{\ln\left[0.076 g g_{\star}^{-1/2}m_{\rm{pl}}m_{\nue}\tilde{\sigma}_0\right] \right\}.
\end{split}
\end{equation}
We have explicitly left in the dependence on today's entropy density
\begin{equation}
s_0=2889.23\,{\rm{cm}}^{-3}\left(\frac{g_{\star s}}{3.9091}\right),
\end{equation}
because the light neutralino and light sneutrino lead to a change in the present day value of \mbox{$g_{\star s}=2+21T_{\nu}/(4 T_{\gamma})$}, through the change in the neutrino-photon temperature ratio (see eq.~\eqref{eq:Tnugamma}).

Figure~\ref{fig:relicabundance} shows the sneutrino mass $m_{\tnu}$ required to achieve the relic abundance $\Omega_{\nue}h^2=0.11$ as a function of the neutralino mass $m_{\nue}$ for three different values of $\sin\theta$. For $\sin\theta\gtrsim0.05$, the required sneutrino mass is $m_{\tnu}\sim\text{few}\times m_{\nue}$. For sufficiently small values of $\sin\theta$, the required sneutrino mass is smaller than the neutralino mass so that the neutralino can no longer be the dark matter. For instance, for $\sin\theta=0.04$, the  sneutrino mass (required to achieve the observed relic abundance) is approximately the same as the neutralino mass at $\sim 19$~MeV. In this article we do not consider the case where the mixed sneutrino is lighter than the neutralino (and thus, potentially, the dark matter particle), although we note that similar bounds deriving from $\Neff$ will also apply in that case.

\begin{figure}[t]
\includegraphics[width=0.99\columnwidth]{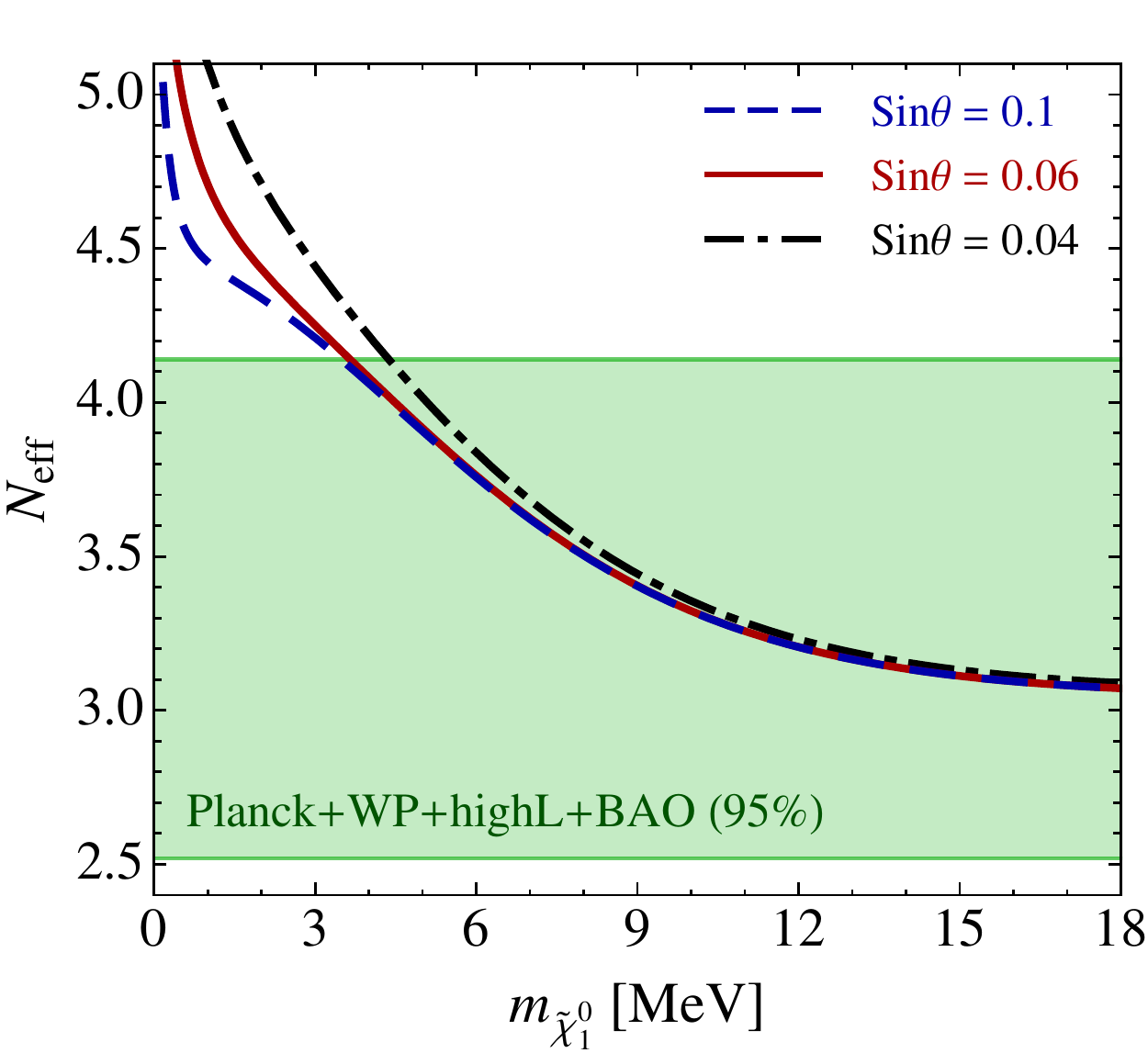}
\caption{The three lines show $\Neff$ as a function of the neutralino mass $m_{\chi}$ for three choices of $\sin\theta$. The sneutrino mass $m_{\tnu}$ has been set to the value in fig.~\ref{fig:relicabundance} so that $\Omega_{\nue}h^2=0.11$. The green shaded area shows the 95\% confidence region $2.52\leq \Neff \leq 4.14$ from the Planck satellite in combination with WMAP-9's polarisation data, SPT's high-$\ell$ data and BAO measurements. We exclude $m_{\nue}\gtrsim3.5$~MeV. The difference in the behaviour of the lines is due to the sneutrino, which contributes when $m_{\tnu}\lesssim10$~MeV.}
\label{fig:bound}
\end{figure}

\subsection{Constraining the model}
In appendix~\ref{app:particleconstraints} we have collected together constraints on this model from invisible widths of the $Z$- and Higgs boson, rare meson decays, collider direct production and beam dump searches as well as constraints from SN1987A, large scale structure and direct detection experiments. These constraints imply that the neutralino must be bino-like and the sneutrino mixing angle \mbox{$\sin\theta\lesssim0.11$}. However, these do not provide us with any constraint on the mass of the neutralino dark matter, apart from the $\mathcal{O}$(keV) mass bound from free streaming and phase space arguments mentioned in the introduction.

Therefore, we now use the $\Neff$ bound from section~\ref{sec:lowerneff} to show that this provides the strongest mass constraint for neutralino dark matter in this model, improving the previous lower bound by three orders of magnitude. Since both the neutralino and sneutrino remain in thermal equilibrium with the neutrinos after they decouple, we can use eq.~\eqref{eq:Neff3} to show
\begin{equation}
N_{\rm{eff}}=3.046 \left[1+\frac{F\left(m_{\nue}/T_{\rm{D}}\right)+F\left(m_{\tnu}/T_{\rm{D}}\right)}{3.046}\right]^{4/3}.
\end{equation}

In fig.~\ref{fig:bound} we plot $\Neff$ as a function of $m_{\nue}$ for the same values of $\sin \theta$ as in fig.~\ref{fig:relicabundance}. The same colour coding is also used in this figure. The sneutrino mass $m_{\tnu}$ has been fixed to the value shown in fig.~\ref{fig:relicabundance} for a given value of $m_{\nue}$ and $\sin\theta$. The shaded green band in fig.~\ref{fig:bound} shows the $95\%$ allowed range of $\Neff$ taken from fig.~\ref{fig:mNeff}, from the Planck analysis including WMAP-9's polarisation data, SPT's high-$\ell$ data and BAO measurements. We find a lower bound $m_{\nue}\gtrsim3.5$~MeV, consistent with the value quoted in eq.~\eqref{eq:MajFermBound} for the case of Majorana fermion dark matter. When the the sneutrino mass drops below $\sim10$~MeV, it can also contribute to $\Neff$. This explains why the three lines behave slightly differently and why the bound when $\sin\theta=0.04$ is slightly weaker.

In order to use the constraint on $\Neff$, we have implicitly assumed that the neutrinos decouple at \mbox{$T_{\rm{D}}=2.3$~MeV} and the neutralino and sneutrino are only in thermal equilibrium with neutrinos. We now show that this is indeed the case.  

The neutralino and sneutrino do not lead to a new interaction between electrons and neutrinos that is stronger than the weak interaction. Therefore, the only change in the decoupling temperature occurs because the expansion rate is slighter faster due to the presence of additional relativistic particles~\cite{2012arXiv1212.1689H}. Neutrinos decouple when $\Gamma\sim G_F^2 T_{\rm{D}}^5=1.66 g_{\star}^{1/2} T_{\rm{D}}^2/m_{\rm{pl}}$. The extra neutralino and sneutrino slightly change $T_{\rm{D}}$ because they contribute to $g_{\star}$. Therefore, the ratio of old and new decoupling temperatures equals the ratio of old and new values of $g_{\star}^{1/6}$. This change is always very small. For instance, taking $m_{\nue}=5$~MeV and $m_{\tnu}=20$~MeV, we find the neutrino decoupling temperature increases to $1.02 \,T_{\rm{D}}$.

We now show that the neutralino and sneutrino are only in thermal equilibrium with neutrinos for temperatures below $T_{\rm{D}}$. In order for this to happen, we have to show that the neutralino and sneutrino are no longer in kinetic equilibrium with the electrons at $T_{\rm{D}}\approx2.3$~MeV. We first consider the neutralino elastically scattering with electrons. The rate of elastic scattering is
\begin{equation}
\Gamma_{\rm{el}}=\langle \sigma_{\nue e} v\rangle n_e\;,
\end{equation}
where $\langle \sigma_{\nue e} v\rangle$ is the thermally averaged elastic scattering cross-section and $n_e$ is the electron number density. The temperature when the last elastic scattering occurs is found from solving $\Gamma_{\rm{el}}=H$, where $H$ is the Hubble expansion rate. However, if the neutralino is non-relativistic, it drops out of kinetic equilibrium before this time because the typical momentum transfer in a single collision, $\Delta p_{\nue}/p_{\nue}\approx \sqrt{3/2}\,T/m_{\nue}$, is small compared to the momentum of the neutralino~\cite{Hofmann:2001bi}. Therefore, $N\approx \sqrt{2/3}\,m_{\nue}/T$ collisions are required to maintain kinetic equilibrium so, when the neutralino is non-relativistic, the temperature of kinetic decoupling is found from solving $\Gamma_{\rm{kd}}=H$, where
\begin{equation}
\Gamma_{\rm{kd}}=\sqrt{\frac{3}{2}}\frac{T}{m_{\nue}}\langle \sigma_{\nue e} v\rangle n_e\;.
\end{equation}

In the limit that the electron mass is zero and left- and right-handed selectrons are degenerate, the thermally averaged elastic scattering cross-section is
\begin{equation}
\langle \sigma_{\nue e} v\rangle=\frac{51}{2 \pi}\left(\frac{m_Z}{m_{\tilde{e}}}\right)^4 G_F^2 \sin^4\theta_W \langle E_e^2\rangle\;,
\end{equation}
where $\langle E_e^2\rangle=12.9 \,T^2$.
With this result, we find that the neutralino drops out of kinetic equilibrium with the electrons when
\begin{equation}
T_{\rm{kd}-e}\approx 2.3 \text{ MeV} \left(\frac{m_{\nue}}{10~\text{MeV}} \right)^{1/4} \left(\frac{m_{\tilde{e}}}{93~\text{GeV}} \right)\;.
\end{equation}
With a more accurate numerical calculation that includes a non-zero electron mass and a thermal average over Fermi-Dirac distributions for the electron and neutralino, we find that $m_{\tilde{e}}>70~\text{GeV}$ for $m_{\nue}=10~\text{MeV}$ in order that $T_{\rm{kd}-e}>2.3~\text{MeV}$. After imposing the ATLAS limit \mbox{$m_{\tilde{e}}>195$~GeV}~\cite{Aad:2012pxa}, we see that the neutralino drops out of kinetic equilibrium with the electrons well before the neutrinos decouple at $T_{\rm{D}}\approx2.3$~MeV.

The sneutrino is kept in equilibrium with the electrons through $Z$- and chargino-exchange. The chargino contribution is smaller than the $Z$-contribution if $m_{\tilde{\chi}_1^{\pm}}>102$~GeV (this assumes the chargino is wino-like to maximise its coupling to the sneutrino, and that the sneutrino has one-third electron flavour). Considering only the $Z$-contribution and ignoring the electron mass we find
\begin{equation}
\langle \sigma_{\tnu e} v\rangle=\frac{\sin^4\theta G_F^2}{\pi}\left(8 \sin\theta_W^4-4 \sin\theta_W^2+1 \right)\langle E_e^2\rangle.
\end{equation}
In this case, the sneutrino drops out of kinetic equilibrium with the electrons when
\begin{equation}
T_{\rm{kd}-e}\approx 2.3 \text{ MeV} \left(\frac{m_{\tnu}}{20~\text{MeV}} \right)^{1/4} \left(\frac{0.37}{\sin\theta} \right)\;.
\end{equation}
Performing a numerical calculation with a non-zero electron mass and a thermal average over Fermi-Dirac distributions for the electron and sneutrino, we also find that $\sin\theta<0.37$ for $m_{\tnu}=20$~MeV. This value of $\sin \theta$ is well above the values we consider, which implies that the sneutrino, like the neutralino, drops out of kinetic equilibrium with the electrons well before the neutrinos decouple.

\section{Conclusions}

In this article we have used the recent Planck measurement of $\Neff$ to set a lower bound on the mass of cold thermal dark matter. These bounds are summarised in eqs.~\eqref{eq:nu_RScalarBound} to~\eqref{eq:nu_DirFermBound} for the case when dark matter is kept in thermal equilibrium with neutrinos and eqs.~\eqref{eq:e_RScalarBound} to~\eqref{eq:e_DirFermBound} for the case when dark matter is kept in thermal equilibrium with electrons and photons. These bounds apply for both $s$- or $p$-wave annihilation cross-sections and are stronger than bounds that can be derived from Big Bang Nucleosynthesis (see table~\ref{tab:bounds}). These bounds apply to models already in the literature such as~\cite{Boehm:2003bt,Ahn:2005ck,Hooper:2008im,Bagnasco:1993st, Pospelov:2000bq,Sigurdson:2004zp,Masso:2009mu, An:2010kc, Fitzpatrick:2010br,2012arXiv1211.0503H} where dark matter annihilates into electrons and could be relevant for models which give a signal at electron recoil direct detection experiments. 

The case where the dark matter remains in equilibrium with neutrinos has received less attention so we have constructed a supersymmetric model to realise this scenario. In this model the dark matter is a light bino-like neutralino and it is kept in equilibrium with neutrinos via a mixed left-right handed sneutrino mediator (see fig.~\ref{fig:spectrum}). In the presence of a such a mediator it is possible to achieve the correct relic density (see fig.~\ref{fig:relicabundance}), leading to changes in $\Neff$ (see fig.~\ref{fig:bound}). The Planck measurement of $\Neff$ leads to a lower bound of 3.5~MeV on the neutralino mass in this scenario.

Our bounds may be evaded if the dark matter's abundance has a non-thermal origin (for example asymmetric dark matter), or if the dark matter is in thermal equilibrium with electrons, neutrinos and photons (in which case there is no change in the standard neutrino-photon temperature ratio). Finally, if the dark matter is in thermal equilibrium with electrons and photons and there is also some `dark radiation' such as sterile neutrinos present, the bounds may be weakened.

\section*{Acknowledgements}

We thank Felix Kahlhoefer, Matthew McCullough, Ian Shoemaker, John March-Russell and Stephen West for discussions. MD and CM also wish to thank Michael Spannowsky for `stimulating' discussions. We thank Gary Steigman for pointing out an error in eq.~\eqref{eq:Neff4} in an early version of this work.

\appendix

\section{Mass bounds with prior on $H_0$}
\label{app:H0}

\begin{figure}[t]
\includegraphics[width=0.99\columnwidth]{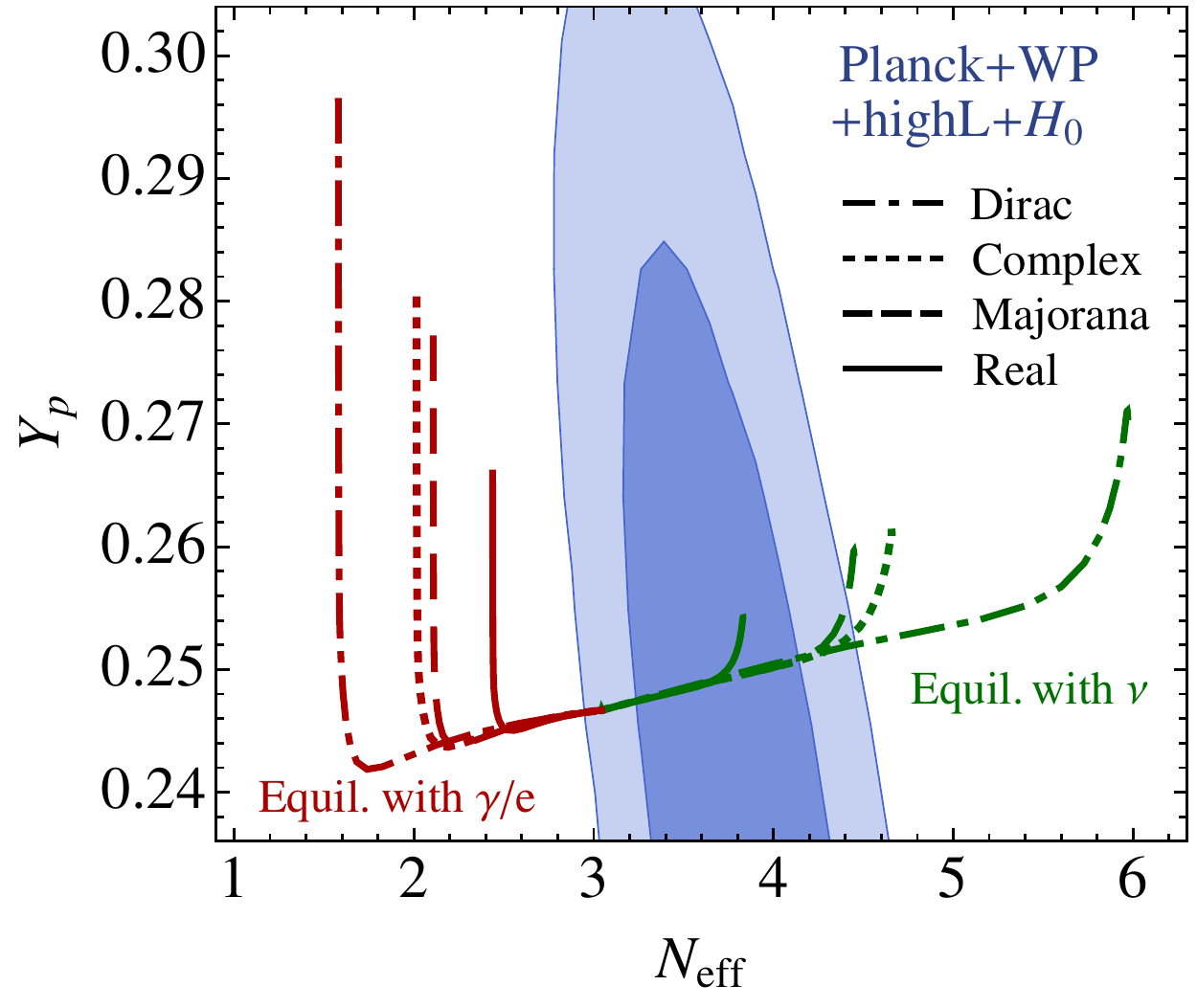}
\caption{The blue regions show the $68\%$ and $95\%$ regions determined from Planck+WP+highL+$H_0$ analysis when both $\Neff$ and $Y_p$ are varied freely. The blue regions are shifted to higher $\Neff$ relative to the contours from the Planck+WP+highL+BAO analysis (c.f.\ fig.~\ref{fig:mNeff}). As before, the green (red) lines indicate the relationship between Yp and Neff for particles in thermal equilibrium with neutrinos (electrons and photons). As $m$ decreases, the prediction for $\Neff$ and $Y_p$ falls outside of the Planck confidence regions.}
\label{fig:neffypH0}
\end{figure}

The value of the Hubble constant $H_0$ determined by Planck is about $2.5\sigma$ lower than the value inferred from direct astrophysical measurements. In particular, Planck find $67.3\pm1.2$~km~s$^{-1}$~Mpc$^{-1}$ while measurements from the Hubble Space Telescope find $73.8\pm2.4$~km~s$^{-1}$~Mpc$^{-1}$~\cite{Riess:2011yx}. This discrepancy could be due to something mundane such as cosmic variance (see e.g.~\cite{Marra:2013rba}) or it could be a sign of physics beyond the simplest $\Lambda$CDM model.

This is relevant to our discussion because $H_0$ is correlated with $\Neff$: larger values of $H_0$ prefer larger values of $\Neff$. In light of this, in this appendix we give the mass bounds derived from Planck when the HST prior from~\cite{Riess:2011yx} is used. The $68\%$ and $95\%$ confidence regions from the Planck+WP+highL+$H_0$ analysis are shown in fig.~\ref{fig:neffypH0}. As expected, the Planck contours are shifted to higher values of $\Neff$.

Requiring that $\Neff$ is consistent with the Planck+WP+highL+$H_0$ result, we exclude the following particle masses at $95\%$~C.L.~for cold thermal dark matter particles in equilibrium with neutrinos:
\begin{align}
&\text{Real scalar}\;&\text{No constraint} \label{eq:nu_RScalarBoundH0}\\
&\text{Complex scalar}\;&m<2.2~\text{MeV}\\ \label{eq:MajFermBoundH0}
&\text{Majorana fermion}\;&m<1.2~\text{MeV}\\ 
&\text{Dirac fermion}\;&m<6.1~\text{MeV} \label{eq:nu_DirFermBoundH0}
\end{align}
Similarly, we exclude the following cold thermal dark matter masses at $95\%$~C.L. when in equilibrium with electrons and photons: 
\begin{align}
&\text{Real scalar}\;&m<6.7~\text{MeV}\label{eq:e_RScalarBoundH0}\\
&\text{Complex scalar}\;&m<13.4~\text{MeV}\\
&\text{Majorana fermion}\;&m<13.4~\text{MeV}\\
&\text{Dirac fermion}\;&m<15.7~\text{MeV}\label{eq:e_DirFermBoundH0}
\end{align}

Since the Planck region is shifted to higher $\Neff$ relative to the region from the Planck+WP+highL+BAO analysis (c.f.\ fig.~\ref{fig:mNeff}), the bounds for particles in equilibrium with neutrinos (electrons and photons) are weaker (stronger) by a few MeV. However, these limits are still stronger than the bounds placed from BBN alone.

\section{Constraints on a light neutralino and a light mixed sneutrino}
\label{app:particleconstraints}
In this section we will show that the scenario we have outlined satisfies current constraints from particle physics experiments and astrophysical and cosmological processes. The light bino case has already been considered extensively in~\cite{Dreiner:2003wh,Dreiner:2009er,Dreiner:2009ic} where it was shown that current constraints are satisfied. We update this analysis to include invisible Higgs decays to neutralinos, and perform a qualitatively similar analysis for the light sneutrino case. We find that all bounds can be evaded. The most stringent bound comes from the invisible Higgs-width which puts a limit of $\sin \theta \lesssim0.11$.

\subsection{$Z$- and $h$-boson widths}\label{app:widths}

The light neutralino and sneutrino couple to the $Z$- and Higgs-bosons. We consider the contribution to the partial width of each in turn. Considering the $Z$ first, the neutralino contribution is
\begin{align}
\Gamma_{Z\to\nue\nue}&=\frac{ G_F m_Z^3}{12 \sqrt{2}\pi} \left(N_{14}^2-N_{13}^2\right)^2 \left(1-\frac{4 m_{\nue}^2}{m_Z^2}\right)^{3/2}\\
&\approx166 \text{ MeV} \times\cos^2 2\beta\left(1-N_{11}^2\right)
\end{align}
while the contribution from one mixed left-and right-handed sneutrino complex scalar is
\begin{align}
\Gamma_{Z\to\tnu\tnu}&=\frac{\sin^4\theta}{24 \pi}\frac{G_F m_Z^3}{\sqrt{2}}\left( 1-\frac{4 m_{\tnu^2}}{m_Z^2}\right)^{3/2}\\
&\approx  83 \text{ MeV}\times  \sin^4\theta\;.
\end{align}
With the limit $\Gamma_{Z\to\rm{inv}}<3$~MeV~\cite{ALEPH:2005ab}, we have that $N_{11}\gtrsim0.87$ and $\sin\theta<0.44$, far from the parameter range we consider.

The invisible width of the Higgs is a notoriously difficult quantity to constrain at hadron colliders as the total width is quite inaccessible. Without some assumption about the Higgs couplings it is not possible to extract a meaningful constraint on this quantity. However, if we assume that the Higgs couplings are as given by the SM and fit to current Higgs data while including an invisible contribution to the width, the invisible branching ratio is constrained to be less than 15-20\%~\cite{Falkowski:2013dza,Giardino:2013bma,Ellis:2013lra,Belanger:2013kya} or equivalently $\Gamma_{h\to\text{inv}}\lesssim0.82$~MeV for an SM width of $4.1$~MeV, as suggested by the LHC Higgs Cross-Section Working Group.
The neutralino contribution to the decay width is
\begin{equation}
\Gamma_{h\to\nue\nue}=\frac{m_h}{16 \pi}\left|Y^{h\nue\nue}\right|^2\left(1-\frac{4 m_{\nue}^2}{m_h^2}\right)^{3/2}
\end{equation}
with \mbox{$Y^{h\nue\nue}=g ( N_{12}-\tan\theta_W N_{11} )(\sin\alpha N_{13}+\cos\alpha N_{14})$}. In the decoupling limit and using $N_{13}/N_{14}\approx -\tan\beta$, valid when the neutralino is very light, we have,
\begin{align}
\Gamma_{h\to\nue\nue}&\approx \frac{G_F m_Z^2 \sin^2\theta_W}{8 \sqrt{2}\pi}  m_h \sin^2 2\beta N_{11}^2 (1-N_{11}^2)\\
&\approx 79 \text{ MeV}\times\sin^2 2\beta\, N_{11}^2 (1-N_{11}^2)
\end{align}
For any value of $N_{11}$, $\tan\beta$ can be chosen so that the contribution to the Higgs width is below the experimental limit. For instance, if $N_{11}>0.9 (0.99)$, the invisible width is below the limit if $\tan\beta>7.6 (2.4)$.

One mixed left- right-handed sneutrino complex scalar contributes
\begin{equation}
\Gamma_{h\to\tnu\tnu}=\frac{Y_{h \tnu \tnu}^2}{16 \pi m_h}\sqrt{1-\frac{4 m_{\tnu}^2}{m_h^2}}\;,
\end{equation}
where
\begin{equation}
Y_{h \tnu \tnu}=\frac{A_1}{2\sqrt{2}}\cos\alpha \sin 2\theta+\frac{m_{W}^2}{\sqrt{2} v}\frac{\sin(\alpha+\beta) \sin^2\theta}{\cos^2\theta_W}\;.
\end{equation}
The first term comes from $A h_u \tilde{\nu} \tilde{n}^*$-term while the second term is the usual contribution of the (left-handed) sneutrino. In the decoupling limit and using the relation \mbox{$A_1 v \sin\beta=\tan\theta(m_{\tilde{\nu}_L}^2-m_{\tnu}^2)\approx\tan\theta m_{\tilde{\nu}_L}^2$}, we have
\begin{align}
Y_{h \tnu \tnu}=\frac{\sin^2\theta m_{\tilde{\nu}_L}^2}{\sqrt{2} v}\left( 1-\left(\frac{m_W}{m_{\tilde{\nu}_L}}\right)^2\frac{\cos 2\beta}{\cos^2\theta_W} \right)\;,
\end{align}
so that
\begin{equation}
\begin{split}
\Gamma_{h\to\tnu\tnu}&\approx 0.82\text{ MeV}\times 
\left(\frac{\sin\theta}{0.11}\right)^4
\left(\frac{m_{\tilde{\nu}_L}}{195 \text{ GeV}}\right)^{4}\\
&\quad \times \left( \frac{1+0.21\left(\frac{195 \text{ GeV}}{m_{\tilde{\nu}_L}}\right)^2\left( \frac{\cos2\beta}{-0.98}\right) }{1.21}\right)^2\;.
\end{split}
\end{equation}
The contribution to the Higgs width is large because of the large $A$-term, required to have reasonably large left- right-mixing (c.f.~eq.~\eqref{eq:mixingangle}). However, we can always choose parameters such that we are below the current experimental limit.

\subsection{Meson decays}

We next consider constraints from partial widths of invisible meson decays. We briefly summarise the neutralino case, previously investigated in~\cite{Dreiner:2009er}, before discussing the implications of their results for the sneutrino. The authors of~\cite{Dreiner:2009er} find that lower bounds cannot be set on the neutralino mass in minimal flavour violation (MFV) scenarios. Only in the case of non-minimal flavour violation can the branching ratios be enhanced, a situation we do not consider in this article. Two-body decays of pseudoscalar mesons to neutralinos proceed at tree-level via an exchanged squark. Two-body decays to sneutrinos are loop-suppressed relative to this, and so pseudoscalar decays do not constrain the sneutrino mass.

One can also consider three-body decays such as $K^- \to \pi^- \tnu \tnu$, where the Standard Model process $K^- \to \pi^- \nu \bar{\nu}$ are already suppressed. These have also been calculated for light neutralinos in~\cite{Dreiner:2009er}, where it was found that the ratio of the SUSY contribution to the observed branching ratio was at most $10^{-6}$. While reproducing the full results of~\cite{Dreiner:2009er} is beyond the scope of the paper, we see no reason why similar results for the sneutrino final state should not also be true.

\subsection{Monojet and monophoton searches}

An active area of recent research has been using collider data to set limits on theories of dark matter. Of particular interest to us are the studies~\cite{Goodman:2010yf,Fox:2011fx}, the first of which deals with monojet limits from the Tevatron and projected limits from LHC, and the second with monophoton searches from LEP. We consider each in turn.

The authors of~\cite{Goodman:2010yf} adopt an effective field theory approach, writing down higher dimensional operators which couple two dark matter fields to two quarks or gluons, suppressed by a scale $M_*$. For light dark matter, the bounds they derive on $M_*$ are $\mathcal{O}(10)$~GeV for Tevatron results, with projected limits of $\mathcal{O}(100)$~GeV at LHC at 14~TeV. As the main mediator in the MSSM for this process are squarks, direct search limits are already far above these bounds. The presence of a light sneutrino does not affect these limits, as processes like $pp\to \tnu \tnu$ are loop suppressed relative to $pp \to \nue \nue$. 

At LEP there are two processes of interest leading to the monophoton final state. These have been considered in~\cite{Fox:2011fx} in an effective field theory context, setting limits on $\Lambda$ the suppression scale of the higher dimensional operator. For the process $e^+ e^- \to \nue \nue \gamma$ this is approximately $\Lambda \sim m_{\tilde{e}} \cos\theta_W /\sqrt{4 \pi \alpha}$ leading to \mbox{$m_{\tilde{e}}\gtrsim 120$~GeV}. The other process which contributes to the monophoton final state is $e^+ e^- \to \tnu\tnu\gamma$ via a t-channel chargino (there is also a $Z$-mediated contribution but this is sub-dominate when the chargino is light). Here the suppression is approximately $\Lambda \sim m_{\tilde{\chi}^{\pm}} \sin\theta_W /(\sqrt{4 \pi \alpha} \sin\theta)$: with $\sin\theta=0.1$ we find $m_{\tilde{\chi}^{\pm}} \gtrsim 21$~GeV. These bounds are easily satisfied.

\subsection{Beam-dump searches}
It has been proposed that high-luminosity fixed-target neutrino experiments can be sensitive to sub-GeV dark matter~\cite{deNiverville:2011it}. In order to assess whether we are constrained by them we adapt the calculation of Ref.~\cite{deNiverville:2012ij} to our case. They investigate sub-GeV dark matter models with both a vector and scalar portal, the scalar portal being the case relevant for us. They find that the sensitivities of MINOS, T2K and MiniBooNE are too weak in the scalar case to set limits (unlike the vector case). In addition to the Higgs mediated diagram contributing to $p\nue \to p \nue$ scattering which they consider, we should also include the contributions from squark exchange. However, as one expects the squark mass to be $m_{\tilde{q}} \sim \mathcal{O}(1)$~TeV this contribution should be suppressed by approximately $m_h^4 / m_{\tilde{q}}^4\sim 10^{-4}$. Our model is thus unconstrained by beam-dump searches.

\subsection{Bounds from SN1987A}\label{app:supernovae}
The neutrinos detected from SN1987A were in good agreement with the theoretical expectation. Requiring that new particles or new interactions do not significantly change the neutrino burst allows bounds to be placed on a wide variety of new particles and interactions (see e.g.~\cite{RaffeltStars}).

The scenario we consider here, namely, of MeV-mass particles with stronger-than-weak interactions with neutrinos and very weak interactions with electrons and nucleons, has not been specifically considered in the literature.
Bounds have previously been set on new particles that have stronger-than-weak interactions with nucleons (or electrons) in addition to stronger-than-weak interactions with neutrinos, leading to a constraint of $m\gtrsim10$~MeV~\cite{Fayet:2006sa}. In our case, the interaction with electrons and nucleons is `weaker-than-weak' so these bounds are not applicable. Our scenario shares more similarity with the Majoron~\cite{Chikashige:1980ui,Gelmini:1980re, Georgi:1981pg}, a Goldstone boson that has large interactions with only the neutrinos, so we adapt bounds placed on that scenario.

We first consider the energy loss argument (see e.g.~\cite{Choi:1989hi,Kachelriess:2000qc,Farzan:2002wx} for the Majoron case). The new particles can act as an efficient energy loss mechanism if they are so weakly interacting that they freely escape from the supernova core, resulting in a shorter Kelvin-Helmholtz cooling phase. Here, we estimate the mean free path of the neutralino to determine if it escapes freely. We have numerically calculated the thermal average of the scattering cross-section (assuming Fermi-Dirac distributions for the neutrino and neutralino with zero chemical potential). For instance, for $m_{\chi}=1-20 \text{ MeV}$, $m_{\tnu}=10-100 \text{ MeV}$ and $\sin \theta=0.1$, we find $\langle \sigma_{\nu \nue } v\rangle\approx 10^{-30} \text{ cm}^2$. This cross-section is large because the sneutrino is on-shell. Assuming a neutrino density $n_{\nu}\approx10^{33} \text{ cm}^{-3}$ and a core temperature $T=10$~MeV~\cite{Sigl:1994da} we find that the mean free path of the neutralino is $\sim 10^{-3}$~cm (this result is not particularly sensitive to the choice of core temperature). 
A similar calculation holds for the sneutrino. Since the mean free path is so much smaller than the geometric dimension of the supernova ($\sim 10$~km), the cooling from neutralinos and sneutrinos is very inefficient and should not decrease the Kelvin-Helmholtz cooling phase~\cite{RaffeltStars}.\footnote{Since the neutralinos do not freely escape the supernova, the SN1987A bounds on the selectron and squark masses from~\cite{Dreiner:2003wh} do not apply.} It has also been argued that the large interaction rate of the neutrinos with thermal neutralinos and sneutrinos increases the cooling time because it takes longer for the neutrinos to diffuse from the core to the neutrino-sphere~\cite{Manohar:1987ec}. However, a more careful analysis treating the neutrinos as a relativistic Fermi gas indicate that this argument is false~\cite{Dicus:1988jh}.

A more robust bound can be placed considering the signal dispersion of $\bar{\nu}_e$. The detection of $\bar{\nu}_e$'s precludes the possibility that they were removed due to `secret interactions' with dark matter particles as they journeyed from the Large Magellanic Cloud (LMC) to Earth~\cite{Kolb:1987qy,Mangano:2006mp}. Integrating the dark matter density along the line of sight between the LMC and the Earth and assuming the Milky Way's dark matter density follows $\rho=\rho_0 (l_0/l)^2$ with $\rho_0=0.3$ GeV cm$^{-3}$ and $l_0=8$ kpc, we can put a bound
\begin{equation}\label{eq:SNbound}
\sigma_{\nu \nue}\lesssim \left(\frac{m_{\nue}}{\rm{MeV}}\right) 10^{-25} \text{ cm}^2
\end{equation} 
on the dark matter scattering cross-section by requiring that that the mean free path for neutrinos is longer than the LMC-Earth distance. 

In the regime where $m_{\tnu}\gtrsim E_{\nu}\gtrsim m_{\nue}$ the elastic scattering cross-section is
\begin{align}
\sigma_{\nu \nue}&=|N_{11}|^4 \sin^4\theta \frac{G_F^2 m_Z^4 \sin^4\theta_W}{3 \pi} \frac{m_{\nue} E_{\nu}}{m_{\tnu}^4}\\
\begin{split}
&\approx \left(\frac{m_{\nue}}{1\text{ MeV}}\right)  10^{-34} \text{ cm}^2\\
&\quad \times  \left( \frac{E_{\nu}}{10 \text{ MeV}}\right) \left( \frac{\sin \theta}{0.1}\right)^4 \left( \frac{20 \text{ MeV}}{m_{\tnu}}\right)^4\;,
\end{split}
\end{align}
well below the limit in eq.~\eqref{eq:SNbound} for the parameter range we consider.

\subsection{Bounds from large scale structure}

Dark matter interactions with Standard Model particles can wash out primordial matter fluctuations and induce a damping of the matter power spectrum at large cosmological scales \cite{2001PhLB..518....8B,Boehm:2003xr,Boehm:2004th}.
In the case of dark matter-neutrino interactions, the magnitude of the effect depends on whether or not the neutrinos have entered their free-streaming regime at the moment of the dark matter thermal decoupling.  One can identify three distinct scenarios \cite{Boehm:2004th}, namely:
scenario A defined by $\Gamma_{\nu-e} > \Gamma_{\nu-\mathrm{dm}}$ and
$\Gamma_{\mathrm{dm}-\nu} < \Gamma_{\nu} \sim \Gamma_{\nu-e}$, scenario B defined by $\Gamma_{\nu-e}  < \Gamma_{\nu-\mathrm{dm}}$ and  $\Gamma_{\mathrm{dm}-\nu} < \Gamma_{\nu} \sim \Gamma_{\nu-\mathrm{dm}}$ and finally scenario C, defined by
$\Gamma_{\mathrm{dm}-\nu} > \Gamma_{\nu}$. We use $\Gamma_{i-j}$ to denote the interaction rate between species $i$ and $j$, and $\Gamma_{\nu}$ to denote the total interaction rate of the neutrinos.

Scenario B assumes a much larger dark matter-neutrino cross-section than
the neutrino-electron elastic scattering cross-section at $T \simeq$ 1 MeV and implies that the neutrino thermal decoupling is actually modified. The effect of such interactions has been studied in detail in \cite{Mangano:2006mp,Serra:2009uu,Fayet:2006sa}. Scenario C requires that the dark matter is coupled to free-streaming neutrinos  and therefore assumes very specific values of the dm-neutrino cross-section, as discussed in \cite{Boehm:2003xr} in the case of MeV dark matter. Finally scenario A corresponds to very small cross-sections (much smaller than the neutrino-electron cross-section at 1 MeV) and the standard value of the temperature for the neutrino thermal decoupling.

One means of parametrising this is by using the opacity $Q=\langle \sigma_{\chi \nu} v \rangle/m_{\chi}$, the thermal average of the neutrino-dark matter elastic cross-section divided by the dark matter mass~\cite{Mangano:2006mp,Serra:2009uu}. The Sloan Digital Sky Survey (SDSS)~\cite{Tegmark:2003uf} constrains this to be $Q \leq 10^{-42} \mbox{ cm}^2 \mbox{ MeV}^{-1}.$
In the extreme non-relativistic limit where $m_{\tnu}> m_{\nue}\gg E_{\nu}$, we find
\begin{align}
Q&= \frac{ 3G_F^2 m_Z^4 \sin^4\theta_W }{2\pi}  \frac{ |N_{11}|^4 \sin^4\theta \langle E_{\nu}^2\rangle}{m_{\nue}(m_{\nue}^2-m_{\tnu}^2)^2}\\
\begin{split}
&\approx 3\times10^{-53}\mbox{ cm}^2 \mbox{ MeV}^{-1} \\
&\quad \times \left( \frac{\sin \theta}{0.1}\right)^4 \left(\frac{1\text{ MeV}}{m_{\nue}}\right)   \left( \frac{20 \text{ MeV}}{m_{\tnu}}\right)^4\;
\end{split}
\end{align}
where we used $\langle E_{\nu} \rangle\approx12.9 \,T_{\nu}^2$ with $T_{\nu}\sim0.75\, T_{\gamma_0}$.
Since this value corresponds to a very small cross-section for MeV neutralinos, our model corresponds to the scenario~A and there is no significant damping of the matter primordial fluctuations.

\subsection{Direct detection}
As is well known, dark matter particles with sub-GeV mass do not have enough kinetic energy to cause a nucleus to recoil with an energy above the low-energy threshold of conventional dark matter direct detection experiments. However, in the sub-GeV mass range, dark matter particles can cause single-electron ionisation signals by scattering off electrons and this signal is detectable~\cite{Essig:2011nj,Graham:2012su}. The first limit on the electron-dark matter scattering cross-section comes from 15~kg-days of data collected by the XENON10 experiment. They constrain the scattering cross-section to be less than $\sigma_e=6\times10^{-36}\text{ cm}^2$ for dark matter of mass 10~MeV.

Considering $\nue+e\to\nue+e$ and assuming degenerate left- and right-handed selectrons, we find
\begin{align}
\sigma_e&=\frac{133}{4 \pi} \left(\frac{m_Z}{m_{\tilde{e}}}\right)^4 G_F^2 \sin^4\theta_W \mu_{e\nue}^2\\
&\approx 3\times10^{-46}\text{ cm}^2 \left(\frac{195 \text{ GeV}}{m_{\tilde{e}}}\right)^4\;,
\end{align}
where $\mu_{e\nue}$ is the neutralino-electron reduced mass. This is far below the XENON10 limit and below the projected limit $\sigma_e\sim10^{-43} \text{ cm}^2$ for a germanium detector with 1~kg-year of data~\cite{Essig:2011nj}. Therefore, it is unlikely that this model will be testable at future direct detection experiments.

\bibliography{ref}

\providecommand{\bysame}{\leavevmode\hbox to3em{\hrulefill}\thinspace}
\begin{thebibliography}{100}

\bibitem{Zeldovich1}
Y.~B. Zel'dovich, Zh. Eksp. Teor. Fiz. \textbf{48} (1965), 986.

\bibitem{Zeldovich2}
Y.~B. Zel'dovich, L.~B. Okun, and S.~B. Pikelner, Usp. Fiz. Nauk. \textbf{84}
  (1965), 113.

\bibitem{Chiu1}
H.~Y. Chiu, Phys. Rev. Lett. \textbf{17} (1966), 712.

\bibitem{Griest:1989wd}
K.~Griest and M.~Kamionkowski, Phys.Rev.Lett. \textbf{64} (1990), 615.

\bibitem{Hut:1977zn}
P.~Hut, Phys.Lett. \textbf{B69} (1977), 85.

\bibitem{Lee:1977ua}
B.~W. Lee and S.~Weinberg, Phys.Rev.Lett. \textbf{39} (1977), 165--168.

\bibitem{Boehm:2003hm}
C.~Boehm and P.~Fayet, Nucl.Phys. \textbf{B683} (2004), 219--263,
  [hep-ph/0305261].

\bibitem{Boyarsky:2008ju}
A.~Boyarsky, O.~Ruchayskiy, and D.~Iakubovskyi, JCAP \textbf{0903} (2009), 005,
   [0808.3902].

\bibitem{Bond:1983hb}
J.~Bond and A.~Szalay, Astrophys.J. \textbf{274} (1983), 443--468.

\bibitem{Padmanabhan:2005es}
N.~Padmanabhan and D.~P. Finkbeiner, Phys.Rev. \textbf{D72} (2005), 023508,
  [astro-ph/0503486].

\bibitem{Mapelli:2006ej}
M.~Mapelli, A.~Ferrara, and E.~Pierpaoli, MNRAS \textbf{369} (2006),
  1719--1724,  [astro-ph/0603237].

\bibitem{Zhang:2006fr}
L.~Zhang, X.-L. Chen, Y.-A. Lei, and Z.-G. Si, Phys.Rev. \textbf{D74} (2006),
  103519,  [astro-ph/0603425].

\bibitem{Galli:2009zc}
S.~Galli, F.~Iocco, G.~Bertone, and A.~Melchiorri, Phys.Rev. \textbf{D80}
  (2009), 023505,  [0905.0003].

\bibitem{Slatyer:2009yq}
T.~R. Slatyer, N.~Padmanabhan, and D.~P. Finkbeiner, Phys.Rev. \textbf{D80}
  (2009), 043526,  [0906.1197].

\bibitem{Hutsi:2011vx}
G.~Hutsi, J.~Chluba, A.~Hektor, and M.~Raidal, Astron.Astrophys. \textbf{535}
  (2011), A26,  [1103.2766].

\bibitem{Galli:2011rz}
S.~Galli, F.~Iocco, G.~Bertone, and A.~Melchiorri, Phys.Rev. \textbf{D84}
  (2011), 027302,  [1106.1528].

\bibitem{Finkbeiner:2011dx}
D.~P. Finkbeiner, S.~Galli, T.~Lin, and T.~R. Slatyer, Phys.Rev. \textbf{D85}
  (2012), 043522,  [1109.6322].

\bibitem{Slatyer:2012yq}
T.~R. Slatyer, Phys.Rev. \textbf{D87} (2013), 123513,  [1211.0283].

\bibitem{Lopez-Honorez:2013cua}
L.~Lopez-Honorez, O.~Mena, S.~Palomares-Ruiz, and A.~C. Vincent, JCAP
  \textbf{1307} (2013), 046,  [1303.5094].

\bibitem{Kolb:1986nf}
E.~W. Kolb, M.~S. Turner, and T.~P. Walker, Phys.Rev. \textbf{D34} (1986),
  2197.

\bibitem{Serpico:2004nm}
P.~D. Serpico and G.~G. Raffelt, Phys.Rev. \textbf{D70} (2004), 043526,
  [astro-ph/0403417].

\bibitem{Beringer:1900zz}
Particle Data Group, J.~Beringer et~al., Phys.Rev. \textbf{D86} (2012), 010001.

\bibitem{2012arXiv1208.0032S}
G.~Steigman, Adv.High Energy Phys. \textbf{2012} (2012), 268321,  [1208.0032].

\bibitem{Boehm:2012gr}
C.~Boehm, M.~J. Dolan, and C.~McCabe, JCAP \textbf{1212} (2012), 027,
  [1207.0497].

\bibitem{Ho:2012ug}
C.~M. Ho and R.~J. Scherrer, Phys.Rev. \textbf{D87} (2013), 023505,
  [1208.4347].

\bibitem{2013arXiv1303.0049S}
G.~Steigman, Phys.Rev. \textbf{D87} (2013), 103517,  [1303.0049].

\bibitem{Ade:2013lta}
Planck Collaboration, P.~Ade et~al.,  (2013),  1303.5076.

\bibitem{Bagnasco:1993st}
J.~Bagnasco, M.~Dine, and S.~D. Thomas, Phys.Lett. \textbf{B320} (1994),
  99--104,  [hep-ph/9310290].

\bibitem{Pospelov:2000bq}
M.~Pospelov and T.~ter Veldhuis, Phys.Lett. \textbf{B480} (2000), 181--186,
  [hep-ph/0003010].

\bibitem{Sigurdson:2004zp}
K.~Sigurdson, M.~Doran, A.~Kurylov, R.~R. Caldwell, and M.~Kamionkowski,
  Phys.Rev. \textbf{D70} (2004), 083501,  [astro-ph/0406355].

\bibitem{Masso:2009mu}
E.~Masso, S.~Mohanty, and S.~Rao, Phys.Rev. \textbf{D80} (2009), 036009,
  [0906.1979].

\bibitem{An:2010kc}
H.~An, S.-L. Chen, R.~N. Mohapatra, S.~Nussinov, and Y.~Zhang, Phys.Rev.
  \textbf{D82} (2010), 023533,  [1004.3296].

\bibitem{Fitzpatrick:2010br}
A.~L. Fitzpatrick and K.~M. Zurek, Phys.Rev. \textbf{D82} (2010), 075004,
  [1007.5325].

\bibitem{2012arXiv1211.0503H}
C.~M. Ho and R.~J. Scherrer, Phys.Lett. \textbf{B722} (2013), 341--346,
  [1211.0503].

\bibitem{Boehm:2003bt}
C.~Boehm, D.~Hooper, J.~Silk, M.~Casse, and J.~Paul, Phys.Rev.Lett. \textbf{92}
  (2004), 101301,  [astro-ph/0309686].

\bibitem{Ahn:2005ck}
K.~Ahn and E.~Komatsu, Phys.Rev. \textbf{D72} (2005), 061301,
  [astro-ph/0506520].

\bibitem{Hooper:2008im}
D.~Hooper and K.~M. Zurek, Phys.Rev. \textbf{D77} (2008), 087302,  [0801.3686].

\bibitem{Essig:2011nj}
R.~Essig, J.~Mardon, and T.~Volansky, Phys.Rev. \textbf{D85} (2012), 076007,
  [1108.5383].

\bibitem{Essig:2012yx}
R.~Essig, A.~Manalaysay, J.~Mardon, P.~Sorensen, and T.~Volansky,
  Phys.Rev.Lett. \textbf{109} (2012), 021301,  [1206.2644].

\bibitem{Graham:2012su}
P.~W. Graham, D.~E. Kaplan, S.~Rajendran, and M.~T. Walters, Phys.Dark Univ.
  \textbf{1} (2012), 32--49,  [1203.2531].

\bibitem{Ricotti:2007au}
M.~Ricotti, J.~P. Ostriker, and K.~J. Mack, Astrophys.J. \textbf{680} (2008),
  829--845,  [0709.0524].

\bibitem{Frampton:2010sw}
P.~H. Frampton, M.~Kawasaki, F.~Takahashi, and T.~T. Yanagida, JCAP
  \textbf{1004} (2010), 023,  [1001.2308].

\bibitem{Hawkins:2011qz}
M.~Hawkins, MNRAS \textbf{415} (2011), 2744--2757,  [1106.3875].

\bibitem{Preskill:1982cy}
J.~Preskill, M.~B. Wise, and F.~Wilczek, Phys.Lett. \textbf{B120} (1983),
  127--132.

\bibitem{Abbott:1982af}
L.~Abbott and P.~Sikivie, Phys.Lett. \textbf{B120} (1983), 133--136.

\bibitem{Dine:1982ah}
M.~Dine and W.~Fischler, Phys.Lett. \textbf{B120} (1983), 137--141.

\bibitem{Cadamuro:2010cz}
D.~Cadamuro, S.~Hannestad, G.~Raffelt, and J.~Redondo, JCAP \textbf{1102}
  (2011), 003,  [1011.3694].

\bibitem{Mangano:2001iu}
G.~Mangano, G.~Miele, S.~Pastor, and M.~Peloso, Phys.Lett. \textbf{B534}
  (2002), 8--16,  [astro-ph/0111408].

\bibitem{Mangano:2005cc}
G.~Mangano, G.~Miele, S.~Pastor, T.~Pinto, O.~Pisanti, et~al., Nucl.Phys.
  \textbf{B729} (2005), 221--234,  [hep-ph/0506164].

\bibitem{Brust:2013ova}
C.~Brust, D.~E. Kaplan, and M.~T. Walters,  (2013),  1303.5379.

\bibitem{DiBari:2013dna}
P.~Di~Bari, S.~F. King, and A.~Merle, Phys.Lett. \textbf{B724} (2013), 77Ð83,
  [1303.6267].

\bibitem{Enqvist:1991gx}
K.~Enqvist, K.~Kainulainen, and V.~Semikoz, Nucl.Phys. \textbf{B374} (1992),
  392--404.

\bibitem{Srednicki:1988ce}
M.~Srednicki, R.~Watkins, and K.~A. Olive, Nucl.Phys. \textbf{B310} (1988),
  693.

\bibitem{Bennett:2012fp}
C.~Bennett, D.~Larson, J.~Weiland, N.~Jarosik, G.~Hinshaw, et~al.,  (2012),
  1212.5225.

\bibitem{Reichardt:2011yv}
C.~Reichardt, L.~Shaw, O.~Zahn, K.~Aird, B.~Benson, et~al., Astrophys.J.
  \textbf{755} (2012), 70,  [1111.0932].

\bibitem{Percival:2009xn}
SDSS Collaboration, W.~J. Percival et~al., MNRAS \textbf{401} (2010),
  2148--2168,  [0907.1660].

\bibitem{Padmanabhan:2012hf}
N.~Padmanabhan, X.~Xu, D.~J. Eisenstein, R.~Scalzo, A.~J. Cuesta, et~al., MNRAS
  \textbf{427} (2012), 2132--2145,  [1202.0090].

\bibitem{Blake:2011en}
C.~Blake, E.~Kazin, F.~Beutler, T.~Davis, D.~Parkinson, et~al., MNRAS
  \textbf{418} (2011), 1707--1724,  [1108.2635].

\bibitem{Anderson:2012sa}
L.~Anderson, E.~Aubourg, S.~Bailey, D.~Bizyaev, M.~Blanton, et~al., MNRAS
  \textbf{428} (2013), 1036--1054,  [1203.6594].

\bibitem{Beutler:2011hx}
F.~Beutler, C.~Blake, M.~Colless, D.~H. Jones, L.~Staveley-Smith, et~al., MNRAS
  \textbf{416} (2011), 3017--3032,  [1106.3366].

\bibitem{Hou:2011ec}
Z.~Hou, R.~Keisler, L.~Knox, M.~Millea, and C.~Reichardt, Phys.Rev.
  \textbf{D87} (2013), 083008,  [1104.2333].

\bibitem{2012arXiv1212.1689H}
C.~M. Ho and R.~J. Scherrer, Phys.Rev. \textbf{D87} (2013), 065016,
  [1212.1689].

\bibitem{Drewes:2013gca}
M.~Drewes, International Journal of Modern Physics E, Vol. \textbf{22} (2013),
  1330019,  [1303.6912].

\bibitem{Auerbach:2001wg}
LSND Collaboration, L.~Auerbach et~al., Phys.Rev. \textbf{D63} (2001), 112001,
  [hep-ex/0101039].

\bibitem{Formaggio:2013kya}
J.~Formaggio and G.~Zeller, Rev. Mod. Phys. 84, \textbf{1307} (2012),
  [1305.7513].

\bibitem{Pisanti:2007hk}
O.~Pisanti, A.~Cirillo, S.~Esposito, F.~Iocco, G.~Mangano, et~al.,
  Comput.Phys.Commun. \textbf{178} (2008), 956--971,  [0705.0290].

\bibitem{Dreiner:2008tw}
H.~K. Dreiner, H.~E. Haber, and S.~P. Martin, Phys.Rept. \textbf{494} (2010),
  1--196,  [0812.1594].

\bibitem{Abdallah:2003xe}
DELPHI Collaboration, J.~Abdallah et~al., Eur.Phys.J. \textbf{C31} (2003),
  421--479,  [hep-ex/0311019].

\bibitem{Bartl:1989ms}
A.~Bartl, H.~Fraas, W.~Majerotto, and N.~Oshimo, Phys.Rev. \textbf{D40} (1989),
  1594.

\bibitem{Choudhury:1999tn}
D.~Choudhury, H.~K. Dreiner, P.~Richardson, and S.~Sarkar, Phys.Rev.
  \textbf{D61} (2000), 095009,  [hep-ph/9911365].

\bibitem{Kachelriess:2000dz}
M.~Kachelriess, JHEP \textbf{0002} (2000), 010,  [hep-ph/0001160].

\bibitem{Gogoladze:2002xp}
I.~Gogoladze, J.~D. Lykken, C.~Macesanu, and S.~Nandi, Phys.Rev. \textbf{D68}
  (2003), 073004,  [hep-ph/0211391].

\bibitem{Dreiner:2003wh}
H.~Dreiner, C.~Hanhart, U.~Langenfeld, and D.~R. Phillips, Phys.Rev.
  \textbf{D68} (2003), 055004,  [hep-ph/0304289].

\bibitem{Dreiner:2009ic}
H.~K. Dreiner, S.~Heinemeyer, O.~Kittel, U.~Langenfeld, A.~M. Weber, et~al.,
  Eur.Phys.J. \textbf{C62} (2009), 547--572,  [0901.3485].

\bibitem{Dreiner:2009er}
H.~Dreiner, S.~Grab, D.~Koschade, M.~Kramer, B.~O'Leary, et~al., Phys.Rev.
  \textbf{D80} (2009), 035018,  [0905.2051].

\bibitem{Dudas:2008eq}
E.~Dudas, S.~Lavignac, and J.~Parmentier, Nucl.Phys. \textbf{B808} (2009),
  237--259,  [0808.0562].

\bibitem{Davies:2011mp}
R.~Davies, J.~March-Russell, and M.~McCullough, JHEP \textbf{1104} (2011), 108,
   [1103.1647].

\bibitem{Hooper:2002nq}
D.~Hooper and T.~Plehn, Phys.Lett. \textbf{B562} (2003), 18--27,
  [hep-ph/0212226].

\bibitem{Bottino:2002ry}
A.~Bottino, N.~Fornengo, and S.~Scopel, Phys.Rev. \textbf{D67} (2003), 063519,
  [hep-ph/0212379].

\bibitem{Belanger:2003wb}
G.~Belanger, F.~Boudjema, A.~Cottrant, A.~Pukhov, and S.~Rosier-Lees, JHEP
  \textbf{0403} (2004), 012,  [hep-ph/0310037].

\bibitem{Pontecorvo:1957qd}
B.~Pontecorvo, Sov.Phys.JETP \textbf{7} (1958), 172--173.

\bibitem{Pontecorvo:1967fh}
B.~Pontecorvo, Sov.Phys.JETP \textbf{26} (1968), 984--988.

\bibitem{Maki:1962mu}
Z.~Maki, M.~Nakagawa, and S.~Sakata, Prog.Theor.Phys. \textbf{28} (1962),
  870--880.

\bibitem{ArkaniHamed:2000bq}
N.~Arkani-Hamed, L.~J. Hall, H.~Murayama, D.~Tucker-Smith, and N.~Weiner,
  Phys.Rev. \textbf{D64} (2001), 115011,  [hep-ph/0006312].

\bibitem{ArkaniHamed:2000kj}
N.~Arkani-Hamed, L.~J. Hall, H.~Murayama, D.~Tucker-Smith, and N.~Weiner,
  (2000),  hep-ph/0007001.

\bibitem{Borzumati:2000mc}
F.~Borzumati and Y.~Nomura, Phys.Rev. \textbf{D64} (2001), 053005,
  [hep-ph/0007018].

\bibitem{Borzumati:2000ya}
F.~Borzumati, K.~Hamaguchi, Y.~Nomura, and T.~Yanagida,  (2000),
  hep-ph/0012118.

\bibitem{Arina:2007tm}
C.~Arina and N.~Fornengo, JHEP \textbf{0711} (2007), 029,  [0709.4477].

\bibitem{MarchRussell:2004uf}
J.~March-Russell and S.~M. West, Phys.Lett. \textbf{B593} (2004), 181--188,
  [hep-ph/0403067].

\bibitem{MarchRussell:2009aq}
J.~March-Russell, C.~McCabe, and M.~McCullough, JHEP \textbf{1003} (2010), 108,
   [0911.4489].

\bibitem{Aad:2012pxa}
ATLAS Collaboration, G.~Aad et~al., Phys.Lett. \textbf{B718} (2013), 879--901,
  [1208.2884].

\bibitem{Buchmueller:2011ab}
O.~Buchmueller, R.~Cavanaugh, A.~De~Roeck, M.~Dolan, J.~Ellis, et~al.,
  Eur.Phys.J. \textbf{C72} (2012), 2020,  [1112.3564].

\bibitem{Buchmueller:2012hv}
O.~Buchmueller, R.~Cavanaugh, M.~Citron, A.~De~Roeck, M.~Dolan, et~al.,
  Eur.Phys.J. \textbf{C72} (2012), 2243,  [1207.7315].

\bibitem{KolbTurner}
E.~Kolb and M.~Turner, \emph{The early universe}, Westview Press, 1990.

\bibitem{Hofmann:2001bi}
S.~Hofmann, D.~J. Schwarz, and H.~Stoecker, Phys.Rev. \textbf{D64} (2001),
  083507,  [astro-ph/0104173].

\bibitem{Riess:2011yx}
A.~G. Riess, L.~Macri, S.~Casertano, H.~Lampeitl, H.~C. Ferguson, et~al.,
  Astrophys.J. \textbf{730} (2011), 119,  [1103.2976].

\bibitem{Marra:2013rba}
V.~Marra, L.~Amendola, I.~Sawicki, and W.~Valkenburg, Phys.Rev.Lett.
  \textbf{110} (2013), 241305,  [1303.3121].

\bibitem{ALEPH:2005ab}
ALEPH Collaboration, DELPHI Collaboration, L3 Collaboration, OPAL
  Collaboration, SLD Collaboration, LEP Electroweak Working Group, SLD
  Electroweak Group, SLD Heavy Flavour Group, S.~Schael et~al., Phys.Rept.
  \textbf{427} (2006), 257--454,  [hep-ex/0509008].

\bibitem{Falkowski:2013dza}
A.~Falkowski, F.~Riva, and A.~Urbano,  (2013),  1303.1812.

\bibitem{Giardino:2013bma}
P.~P. Giardino, K.~Kannike, I.~Masina, M.~Raidal, and A.~Strumia,  (2013),
  1303.3570.

\bibitem{Ellis:2013lra}
J.~Ellis and T.~You, JHEP \textbf{1306} (2013), 103,  [1303.3879].

\bibitem{Belanger:2013kya}
G.~Belanger, B.~Dumont, U.~Ellwanger, J.~Gunion, and S.~Kraml, Phys.Lett.
  \textbf{B723} (2013), 340--347,  [1302.5694].

\bibitem{Goodman:2010yf}
J.~Goodman, M.~Ibe, A.~Rajaraman, W.~Shepherd, T.~M. Tait, et~al., Phys.Lett.
  \textbf{B695} (2011), 185--188,  [1005.1286].

\bibitem{Fox:2011fx}
P.~J. Fox, R.~Harnik, J.~Kopp, and Y.~Tsai, Phys.Rev. \textbf{D84} (2011),
  014028,  [1103.0240].

\bibitem{deNiverville:2011it}
P.~deNiverville, M.~Pospelov, and A.~Ritz, Phys.Rev. \textbf{D84} (2011),
  075020,  [1107.4580].

\bibitem{deNiverville:2012ij}
P.~deNiverville, D.~McKeen, and A.~Ritz, Phys.Rev. \textbf{D86} (2012), 035022,
   [1205.3499].

\bibitem{RaffeltStars}
G.~G. Raffelt, \emph{Stars as laboratories for fundamental physics}, The
  University of Chicago Press, 1996.

\bibitem{Fayet:2006sa}
P.~Fayet, D.~Hooper, and G.~Sigl, Phys.Rev.Lett. \textbf{96} (2006), 211302,
  [hep-ph/0602169].

\bibitem{Chikashige:1980ui}
Y.~Chikashige, R.~N. Mohapatra, and R.~Peccei, Phys.Lett. \textbf{B98} (1981),
  265.

\bibitem{Gelmini:1980re}
G.~Gelmini and M.~Roncadelli, Phys.Lett. \textbf{B99} (1981), 411.

\bibitem{Georgi:1981pg}
H.~M. Georgi, S.~L. Glashow, and S.~Nussinov, Nucl.Phys. \textbf{B193} (1981),
  297.

\bibitem{Choi:1989hi}
K.~Choi and A.~Santamaria, Phys.Rev. \textbf{D42} (1990), 293--306.

\bibitem{Kachelriess:2000qc}
M.~Kachelriess, R.~Tomas, and J.~Valle, Phys.Rev. \textbf{D62} (2000), 023004,
  [hep-ph/0001039].

\bibitem{Farzan:2002wx}
Y.~Farzan, Phys.Rev. \textbf{D67} (2003), 073015,  [hep-ph/0211375].

\bibitem{Sigl:1994da}
G.~Sigl and M.~S. Turner, Phys.Rev. \textbf{D51} (1995), 1499--1509,
  [astro-ph/9405036].

\bibitem{Manohar:1987ec}
A.~Manohar, Phys.Lett. \textbf{B192} (1987), 217.

\bibitem{Dicus:1988jh}
D.~A. Dicus, S.~Nussinov, P.~B. Pal, and V.~L. Teplitz, Phys.Lett.
  \textbf{B218} (1989), 84.

\bibitem{Kolb:1987qy}
E.~W. Kolb and M.~S. Turner, Phys.Rev. \textbf{D36} (1987), 2895.

\bibitem{Mangano:2006mp}
G.~Mangano, A.~Melchiorri, P.~Serra, A.~Cooray, and M.~Kamionkowski, Phys.Rev.
  \textbf{D74} (2006), 043517,  [astro-ph/0606190].

\bibitem{2001PhLB..518....8B}
C.~{B{\oe}hm}, P.~{Fayet}, and R.~{Schaeffer}, Physics Letters B \textbf{518}
  (2001), 8--14,  [astro-ph/0012504].

\bibitem{Boehm:2003xr}
C.~Boehm, H.~Mathis, J.~Devriendt, and J.~Silk, MNRAS \textbf{360(1)} (2005),
  282--287,  [astro-ph/0309652].

\bibitem{Boehm:2004th}
C.~Boehm and R.~Schaeffer, Astron.Astrophys. \textbf{438, Issue 2} (August I
  2005), 419--442,  [astro-ph/0410591].

\bibitem{Serra:2009uu}
P.~Serra, F.~Zalamea, A.~Cooray, G.~Mangano, and A.~Melchiorri, Phys.Rev.
  \textbf{D81} (2010), 043507,  [0911.4411].

\bibitem{Tegmark:2003uf}
SDSS Collaboration, M.~Tegmark et~al., Astrophys.J. \textbf{606} (2004),
  702--740,  [astro-ph/0310725].

\end{thebibliography}
\bibliographystyle{ArXiv}

\end{document}